\documentclass[11pt]{article}
\usepackage{amsmath, amssymb, graphicx}
\usepackage{booktabs}
\usepackage{subcaption}
\usepackage{siunitx}
\usepackage{dirtree}
\usepackage{xurl}
\usepackage[round]{natbib}
\title{A Benchmark Dataset for Satellite-Based Estimation and Detection of Rain}
\author{%
	  Simon Pfreundschuh\thanks{Corresponding author: simon.pfreundschuh@colostate.edu} \textsuperscript{,1}, %
	Malarvizhi Arulraj\textsuperscript{2}, %
	Ali Behrangi\textsuperscript{3},\\ %
	Linda Bogerd\textsuperscript{2, 7},%
	Alan James Peixoto Calheiros\textsuperscript{4}, %
	Daniele Casella\textsuperscript{5},\\ %
	Neda Dolatabadi\textsuperscript{6}, %
	Clement Guilloteau\textsuperscript{6}, %
	Jie Gong\textsuperscript{7},\\ %
	Christian D. Kummerow\textsuperscript{1}, %
	Pierre Kirstetter\textsuperscript{8},  %
	Gyuwon Lee\textsuperscript{9},\\ %
	Maximilian Maahn\textsuperscript{10}, %
	Lisa Milani\textsuperscript{2, 7}, %
  Giulia Panegrossi\textsuperscript{5} \\ %
	Rayana Palharini\textsuperscript{11},%
	Veljko Petković\textsuperscript{2, 12}, %
	Soorok Ryu\textsuperscript{9},\\ %
	Paolo Sanò\textsuperscript{5}, %
	Jackson Tan\textsuperscript{7, 13}
}
\date{
  \begin{flushleft}
	\footnotesize
	\textsuperscript{1} Department of Atmospheric Science, Colorado State University \\
	\textsuperscript{2} Earth System Science Interdisciplinary Center, University of Maryland  \\
	\textsuperscript{3} Department of Hydrology and Atmospheric Sciences, University of Arizona\\
	\textsuperscript{4} Instituto Nacional de Pesquisas Espaciais \\
	\textsuperscript{5} Institute of Atmospheric Sciences and Climate, Italian National Research Council \\
	\textsuperscript{6} Department of Civil and Environmental Engineering, University of California Irvine \\
	\textsuperscript{7} NASA Goddard Space Flight Center \\
  \textsuperscript{8} School of Meteorology \& School of Civil Engineering and Environmental Science, University of Oklahoma \\
	\textsuperscript{9} Department of Atmospheric Sciences, Kyungpook National University \\
	\textsuperscript{10} Institute for Meteorology, Leipzig University \\
  \textsuperscript{11} Departamento de Prevención de Riegos y Medio Ambiente, Universidad Tecnológica Metropolitana, \\
	\textsuperscript{12} Cooperative Institute for Satellite Earth System Studies, University of Maryland  \\
  \textsuperscript{13} University of Maryland, Baltimore County \\
	\end{flushleft}
}
\begin{document}
\maketitle

\begin{abstract}

	Accurately tracking the global distribution and evolution of precipitation is essential for both research and operational meteorology. Satellite observations remain the only means of achieving consistent, global-scale precipitation monitoring. While machine learning has long been applied to satellite-based precipitation retrieval, the absence of a standardized benchmark dataset has hindered fair comparisons between methods and limited progress in algorithm development.

	To address this gap, the International Precipitation Working Group has developed SatRain, the first AI-ready benchmark dataset for satellite-based detection and estimation of rain, snow, graupel, and hail. SatRain includes multi-sensor satellite observations representative of the major platforms currently used in precipitation remote sensing, paired with high-quality reference estimates from ground-based radars corrected using rain gauge measurements. It offers a standardized evaluation protocol to enable robust and reproducible comparisons across machine learning approaches.

	In addition to supporting algorithm evaluation, the diversity of sensors and inclusion of time-resolved geostationary observations make SatRain a valuable foundation for developing next-generation AI models to deliver more accurate, detailed, and globally consistent precipitation estimates.
\end{abstract}

\section{Background and Summary}

Precipitation, the deposition of water in liquid or frozen form from the
atmosphere onto the Earth's surface, is essential for sustaining ecosystems and
a wide range of human activity. However, extreme events at both ends of the
climatological distribution of precipitation, such as droughts or heavy
precipitation, can cause substantial damage to societies and human livelihoods.
Monitoring the global distribution of precipitation is therefore critical not
only for advancing scientific understanding of the processes that shape
precipitation patterns and drive extreme events but also economic planning and
civil security. Despite its crucial role in many aspects of economic and social
life on Earth, precipitation estimation still faces significant challenges in
meeting the needs of hydrological and climate research, as well as operational
applications. Precipitation is one of the most difficult atmospheric parameters
to measure accurately because its estimation from both satellite and ground-based
observations is complicated by several factors: its high spatial and temporal
variability; its phase (liquid, solid, or mixed); its microphysical compositions
(including particle shape, densities, and sizes); and the difficulties involved
in converting radiometric measurements into quantitative precipitation
estimates \citep{Levizzani2020Satellite}.

Owing to its significance for socio-economic activities and its fundamental role
within the climate system, numerous techniques have been developed to monitor
and quantify it. The three principal approaches are rain gauges, ground-based
radar, and satellite observations. Rain gauges yield direct and highly accurate
measurements but are largely confined to continental regions and are often
irregularly distributed \citep{Kidd2017RainGauges}. In addition, because gauges
measure precipitation only at a single point, they cannot adequately represent
the spatial structure of rainfall systems. Ground-based weather radars yield
spatially-continuous estimates at high spatial resolution but their coverage
remains geographically limited. In contrast, satellites offer consistent,
near-global observations, making them the only means of obtaining continuous and
spatially comprehensive estimates of precipitation.

However, the accuracy with which precipitation can be estimated or detected from
satellite observations varies significantly with sensor type and observing
conditions \citep{Stephens2007RemoteSensing}. Although a small number of
precipitation radars have been deployed to measure precipitation from space,
their spatial and temporal coverages are severly limited. Therefore, global
precipitation monitoring has to rely mostly on passive sensors. Passive
microwave (PMW) sensors operating in the 10-89 GHz range can detect emission signals
associated with precipitation particles over the ocean. However, over land the
emission signal from precipitation is hardly distinguishable from the high and
variable emission signal from the surface. While higher microwave frequencies
($> \SI{80}{\giga \hertz}$) are less sensitive to surface properties, their
information content primarily derives from scattering by large rain drops and
ice particles \citep{Bennartz2003Sensitivity}, providing a weaker link to the
precipitation at the surface. Furthermore, the spatial resolution achievable
with passive microwave sensors is limited, requiring deployment in low-Earth
orbits. As a result, even for satellite missions comprising constellations of
multiple sensors, such as the Global Precipitation Measurement (GPM) mission
\citep{Hou2014GPMMission}, the revisit times can exceed three hours in the
tropics (30 °S - 30 °N).

By contrast, geostationary satellites offer near-continuous temporal coverage
over much of the globe, with spatial resolutions on the order of a few
kilometers. Their main limitation is that they operate only in the visible and
infrared (Vis/IR) bands, which are primarily sensitive to the cloud tops and therefore
provide a less direct link to surface precipitation than passive microwave
observations.

\begin{figure}[htbp]
	\centering \includegraphics[width=1.0\textwidth]{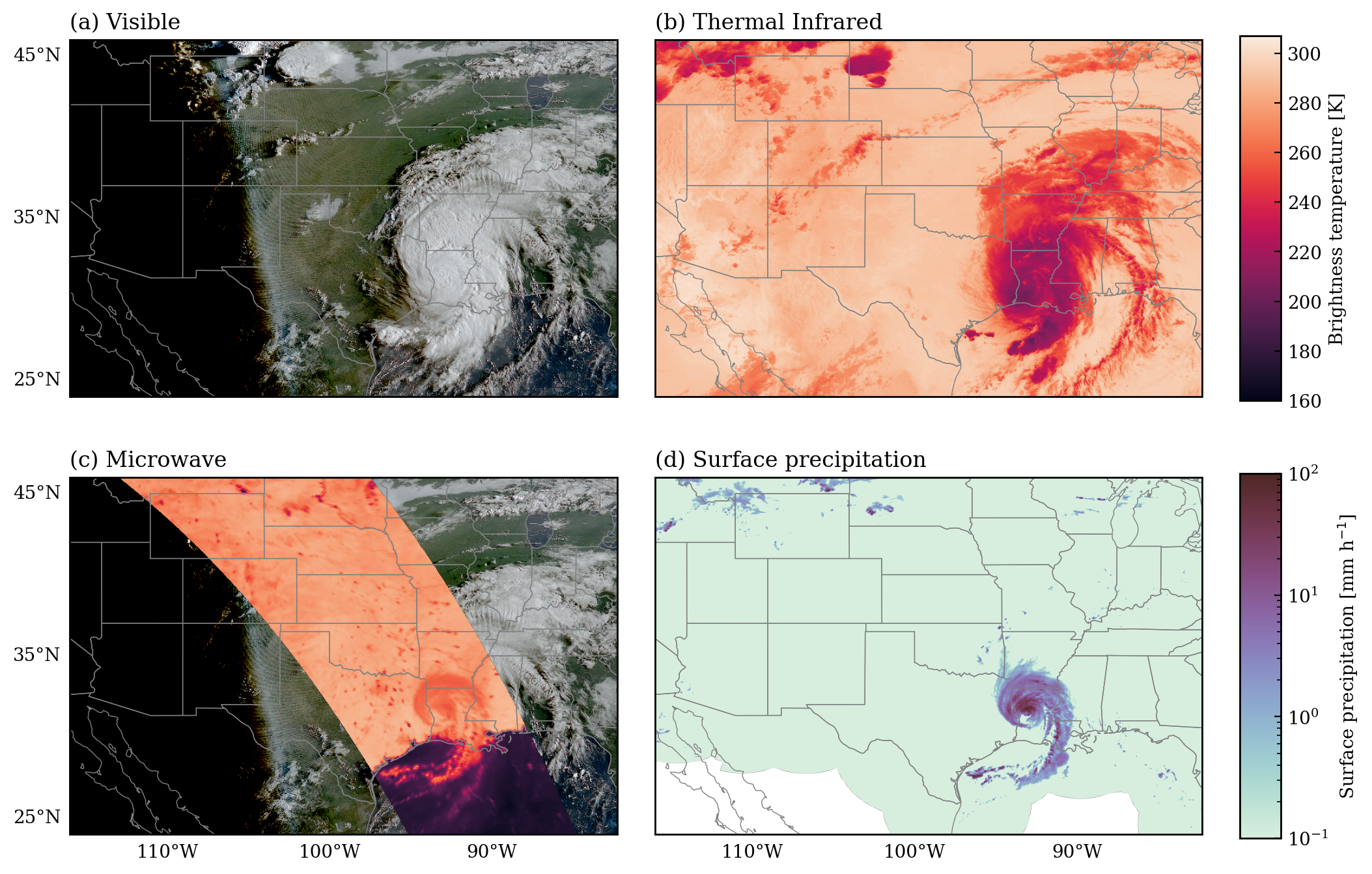}
	\caption{Satellite observations and surface precipitation estimates from the
		landfall of Hurricane Laura on August 27, 2020, at 12:41 UTC. Panel (a)
		shows a true-color composite from the Advanced Baseline Imager (ABI) aboard
		the GOES-16 geostationary satellite. Panel (b) displays geostationary
		thermal infrared imagery at $\SI{11}{\micro \meter}$. Panel (c) presents
		passive microwave observations from the GPM Microwave Imager (GMI) 36.5-GHz,
		horizontally-polarized channel, overlaid on the GOES true-color RGB. Panel
		(d) depicts surface precipitation rate estimates from NOAA’s Multi-Radar
		Multi-Sensor product.}
	\label{fig:observations_conus}
\end{figure}

Figure~\ref{fig:observations_conus} illustrates the key characteristics of the
various types of satellite observations used for the remote sensing of
precipitation. Panel (a) presents a true-color composite from the Advanced
Baseline Imager (ABI, \citeauthor{Schmit2005Introducing},
\citeyear{Schmit2005Introducing}) aboard GOES-16, showing Hurricane Laura as an
expansive cloud system in the southeastern portion of the domain. Visible
imagery such as this true-color composite can reveal detailed cloud structures
but is limited to daylight hours due to its reliance on reflected sunlight.
Panel (b) displays thermal infrared (IR) imagery with a wavelength of
\SI{11}{\micro \meter} observed from a geostationary platform. At this
wavelength, the clear-sky atmosphere exhibits high transmittance, while clouds
are opaque. Consequently, the measured radiances primarily originate from the
Earth’s surface in cloud-free regions and from cloud tops in cloudy regions. Due
to the vertical thermal structure of the atmosphere, the cloud tops appear as
areas of cold brightness temperatures against a relatively warmer background.
Panel (c) shows horizontally-polarized passive microwave observations at a
frequency of $\SI{36.5}{\giga \hertz}$, corresponding to a wavelength of around
$\SI{8}{\milli \meter}$. Surface-sensitive passive microwave observations are
characterized by a strong contrast between ocean and land surfaces. Over the
radiatively cold ocean background, Hurricane Laura's rainband appears as a
region of enhanced brightness temperature due to emission from liquid
hydrometeors. Over land, the surface itself emits strongly at microwave
frequencies, making it difficult to isolate the emission signal from raindrops.
Instead, precipitation is primarily detected through the scattering signature of
large raindrops and ice particles, which reduce the observed brightness
temperatures by attenuating surface emission.

Comparing the three panels to the corresponding radar-based surface
precipitation estimates underscores the strength of passive microwave
observations: while visible and IR imagery primarily depict cloud-top features
that may not correlate with surface rainfall, microwave observations exhibit
higher spatial coherence with surface precipitation. The principal limitation of
passive microwave imagery, however, is its limited spatiotemporal coverage due
to narrower swath widths, longer revisit times, and lower spatial resolution.

\subsection{From Satellite Observations to Precipitation Estimates}

Satellite observations only contain an indirect signal from the near-surface
precipitation and thus require careful processing to produce reliable surface
precipitation estimates. This conversion of satellite measurements into
precipitation estimates is commonly referred to as retrieval. While a
physics-based formulation of precipitation retrieval algorithms is possible,
practical implementations typically require a range of simplifications and
ad-hoc assumptions, such as normally distributed retrieval targets and
measurement errors or close-to linear relationships between atmospheric
variables and satellite observations \citep{Boukabara2011MiRS,
	Kummerow2015GPROF, Maahn2020OptimalEstimation}. Because of these difficulties,
purely empirical approaches have long been used to directly relate satellite
observations to precipitation estimates obtained from other measurement
techniques \citep{ Griffith1978RainEstimation, Adler1988InfraredRainfall,
	Hong2004Precipitation}. The rise of machine learning (ML) and more recent AI
techniques have led to the development of a range of new ML-based
algorithms that yield promising results \citep{Sadeghi2019Persiann,
	Pfreundschuh2022BrazilRetrieval, Pfreundschuh2022Gprof, Gorooh2023LEOGEO}.

\subsection{The Need for a Unified Benchmark Dataset}

The number of ML-based satellite precipitation retrievals is
growing \citep{Sano2018PNPR, Amell2025Probabilistic}, however the algorithms
described in the literature are difficult to compare. This is largely because
they are typically developed for specific sensors, regions, time periods, and
even resolutions. Given the high spatiotemporal variability of precipitation,
such differences in sampling and geographic focus significantly influence
accuracy metrics, rendering published results incomparable. Moreover, the
performance of empirical retrieval algorithms is influenced by the volume,
quality, and spatiotemporal sampling of the training and evaluation data, and
thus does not solely reflect the intrinsic qualities of a specific algorithm or
model.

This lack of comparability makes it difficult to isolate algorithmic
improvements driven by advancements in ML from differences
introduced by the choice of training and evaluation data. To address this
challenge, the International Precipitation Working Group (IPWG), a permanent
Working Group of the Coordination Group for Meteorological Satellites (CGMS),
has established a ML working group tasked with developing a
standardized benchmarking dataset for empirical and ML-based
precipitation retrievals \citep{Kubota2025IPWG}. The result of this effort is
the Satellite-Based Estimation and Detection of Rain (SatRain) dataset, an
AI-ready, large-scale benchmark dataset for the development and evaluation of
precipitation retrieval algorithms covering a wide range of observations
modalities and multiple climate zones. Despite its name, the dataset is not
limited to rain but includes all types of precipitation encountered during the
training and testing periods thus providing a comprehensive resource for
developing and testing precipitation retrieval algorithms.

\subsection{The SatRain Dataset}

The SatRain dataset integrates multi-sensor satellite observations with
gauge-corrected, ground-based radar precipitation estimates. It provides a
large, curated training set over the continental United States (CONUS),
encompassing diverse climate regimes ranging from subtropical humid regions to
arid deserts, mountainous terrain, and temperate to cold continental zones. Data
are available both on a $\SI{0.036}{\degree}$ regular latitude–longitude grid and the native
sampling of the passive microwave sensors. All input and reference fields are
consistently mapped to these two spatial representations, enabling direct use
for both pixel-based and image-based AI algorithms and ensuring the dataset’s
AI-readiness. The satellite observations span a wide range of sensing modalities
relevant to precipitation remote sensing, including temporally resolved imagery
from geostationary platforms. To support model generalization studies, SatRain
also includes independent test sets from Korea and Austria, covering distinct
climate regimes and incorporating alternative reference measurement techniques.

\section{Methods}

\subsection{Data Sources}

The SatRain dataset integrates four principal data sources:

\begin{enumerate}
	\item Passive microwave observations,
	\item Visible and infrared observations from geostationary platforms,
	\item ancillary environmental data,
	\item reference precipitation estimates from ground-based weather radars  over CONUS,
	\item independent precipitation estimates from ground-based radars over Korea and gauge measurements in Austria.
\end{enumerate}

A key challenge in creating the SatRain dataset was reconciling differences in
spatial resolution and sampling across sensors. To balance flexibility with
manageable dataset size, SatRain data is provided in two spatial sampling geometries.
The first geometry is a regular latitude-longitude grid at
$\SI{0.036}{\degree}$ resolution. This resolution matches the native grid of a
major global geostationary IR dataset \citep{NCEP_CPC_L3_IR} and remains finer
than the effective resolution of current satellite precipitation products.
The second is the \textit{on-swath} geometry, which retains the native spatial sampling of
the PMW base sensor. Because many precipitation retrieval algorithms were
originally designed to operate on the native sensor sampling, SatRain retains this
format to support these retrieval designs and cross-comparison against existing
operational retrievals such as the Goddard Profiling Algorithm
\citep{Kummerow2015GPROF}.

Since PMW observations are only available at discrete overpass times for a given
location, SatRain is organized around these overpasses, with coincident data
from other sources collocated accordingly. As a result, the available training
and testing samples are limited to the overpass times of the underlying PMW
sensors.

\subsubsection{Passive Microwave Observations}

\begin{figure}[htbp] 
	\centering
	\includegraphics[width=1.0\textwidth]{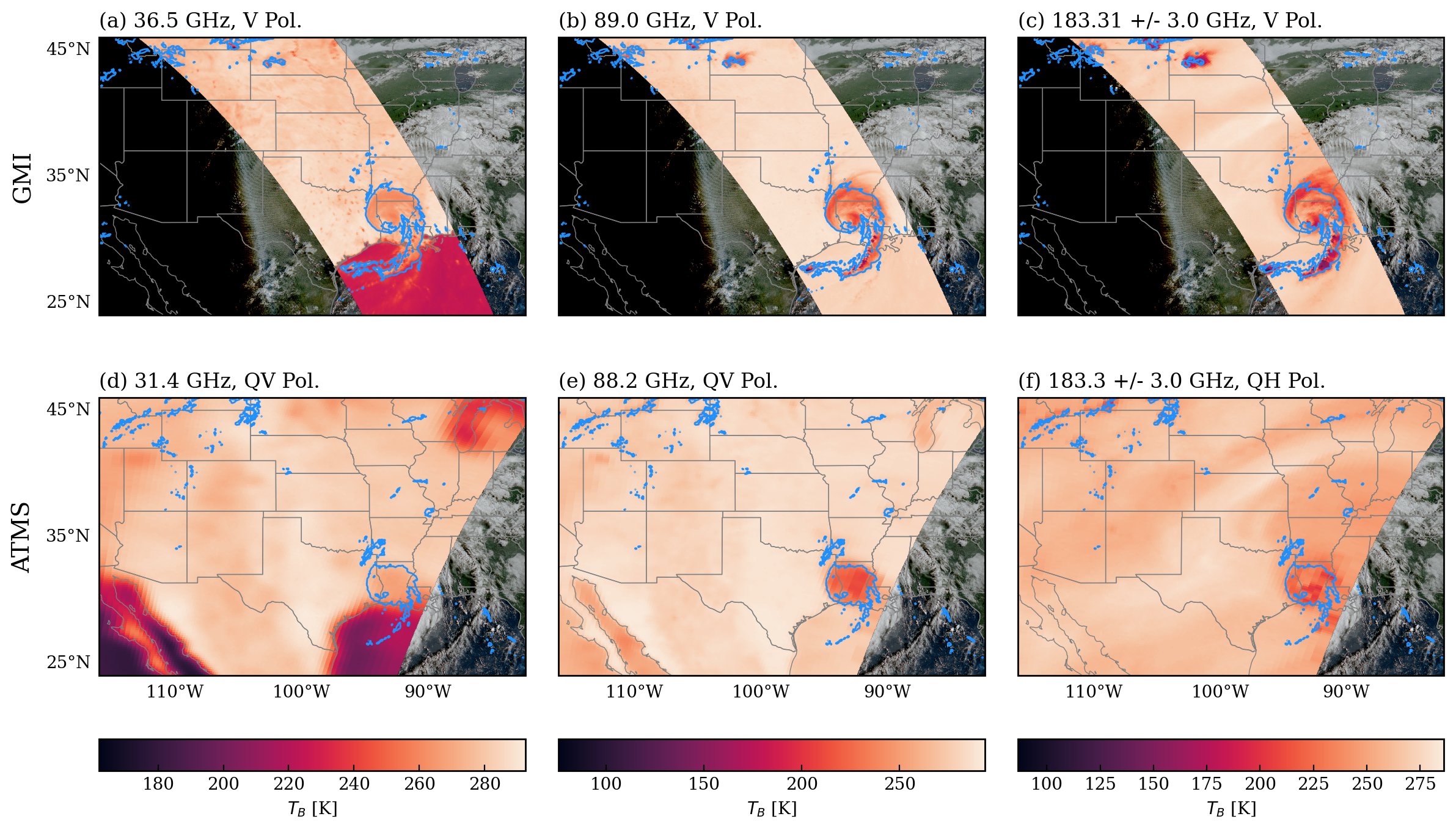}
	\caption{
		Passive microwave observations from the GMI (a - c) and ATMS (d - f)
		sensors from the landfall of Hurricane Laura on August 27, 2020. The top
		row displays three selected channels from the GMI sensor, highlighting
		its high-resolution but limited swath. The bottom row presents
		corresponding channels from the ATMS sensor for comparison. The blue
		contour line encloses regions with surface precipitation rates exceeding
		1 mm h$^{-1}$.
	}
	\label{fig:observations_pmw}
\end{figure}

The PMW observations used to build the SatRain dataset are sourced from two
different sensors: the GPM Microwave Imager (GMI, \citeauthor{Draper2015GMI},
\citeyear{Draper2015GMI}) aboard the GPM Core Observatory, and a selection of
channels of the Advanced Technology Microwave Sounder (ATMS,
\citeauthor{Goldberg2006ATMS}, \citeyear{Goldberg2006ATMS}) aboard the NOAA-20
satellite \citep{Goldberg2013Joint}. The GMI and ATMS sensors were chosen to
represent two ends of the spectrum of PMW instrumentation used for measuring
precipitation. GMI, the flagship PMW sensor of the GPM constellation, has been
designed for precipitation remote sensing and features optimized spectral
coverage and comparably high spatial resolution. In contrast, ATMS was developed
primarily for operational weather forecasting and, compared to GMI, offers fewer
precipitation-sensitive channels and significantly coarser spatial resolution.

sFigure~\ref{fig:observations_pmw} compares GMI and ATMS observations of
Hurricane Laura, collected at 12:41 UTC and 08:41 UTC, respectively, on August
27, 2020. GMI offers higher spatial resolution than ATMS, allowing finer cloud
and precipitation structures to be resolved, but this advantage comes at the
cost of a narrower swath and thus reduced spatial coverage. The instruments also
differ in their scanning strategies. GMI operates as a conical scanner, with its
antenna beam sweeping out a cone relative to the spacecraft, producing
near-circular footprints at a nearly constant Earth-incidence angle. This
geometry ensures uniform polarization and viewing characteristics across the
swath. ATMS, by contrast, employs a cross-track scanning approach, in which the
antenna beam sweeps perpendicular to the spacecraft track. For cross-track
sensors, footprint size increases with scan angle, resulting in coarser
resolution toward the swath edges, while variations in path length and
polarization angle introduce additional systematic differences across the
scan.ystematic differences across the scan.

The PMW observations in SatRain are taken from the calibrated brightness
temperature products of the GPM mission for GMI \citep{Berg2022_GMI_L1C_R_V07}
and from ATMS aboard the NOAA-20 satellite \citep{Berg2022_ATMS_NOAA20_1C_V07}.
The channels used in the dataset are summarized in Table~\ref{tab:pmw_channels},
along with their polarizations and approximate footprint sizes. Because the GPM
data product does not include the ATMS temperature-sounding channels near
\SI{50}{\giga\hertz}, these channels are excluded from the SatRain ATMS
subset.

\begin{table}[htbp]
	\centering
	\begin{subtable}{0.45\textwidth}
		\centering
		\begin{tabular}{ccc}
			\multicolumn{3}{c}{GMI}           \\[0.5ex]
			\toprule
			Freq. [GHz] & Pol. & FP [km x km] \\
			\midrule
			10.65       & V    & 32 x 19      \\
			10.65       & H    & 32 x 19      \\
			18.7        & V    & 18 x 11      \\
			18.7        & H    & 18 x 11      \\
			23.8        & V    & 15 x 9.2     \\
			36.5        & V    & 14 x 8.6     \\
			36.5        & H    & 14 x 8.6     \\
			89.0        & V    & 7.2 x 4.4    \\
			89.0        & H    & 7.2 x 4.4    \\
			166.5       & V    & 7.2 x 4.4    \\
			166.5       & H    & 7.2 x 4.4    \\
			\bottomrule
		\end{tabular}
	\end{subtable}%
	\hfill%
	\begin{subtable}{0.45\textwidth}
		\centering
		\begin{tabular}{ccc}
			\multicolumn{2}{c}{ATMS}          \\[0.5ex]
			\toprule
			Freq. [GHz]      & Pol. & FP [km] \\
			\midrule
			23.800           & QV   & 75      \\
			31.400           & QV   & 75      \\
			88.2             & QV   & 32      \\
			165.5            & QH   & 16      \\
			183.31 $\pm$ 7.0 & QH   & 16      \\
			183.31 $\pm$ 4.5 & QH   & 16      \\
			183.31 $\pm$ 3.0 & QH   & 16      \\
			183.31 $\pm$ 1.8 & QH   & 16      \\
			183.31 $\pm$ 1.0 & QH   & 16      \\
			\bottomrule
		\end{tabular}
	\end{subtable}
	\caption{
		Frequencies (Freq.), polarizations (Pol.), and footprint (FP) sizes of the
		PMW observations included in the SatRain dataset. For GMI, footprint sizes are
		reported as the full width at half maximum (FWHM) along and across the
		boresight direction. For ATMS, only the nadir FWHM is given; at nadir the
		footprint is circular and therefore represented by a single value. The
		polarization of ATMS are denoted quasi-horizontal (QH) and quasi-vertical (QV)
		as the polarization mixture changes with the earth-incidence angle.
	}
	\label{tab:pmw_channels}
\end{table}

\subsubsection{Geostationary Visible and Infrared Observations}

\begin{figure}[htbp] 
	\centering
	\includegraphics[width=1.0\textwidth]{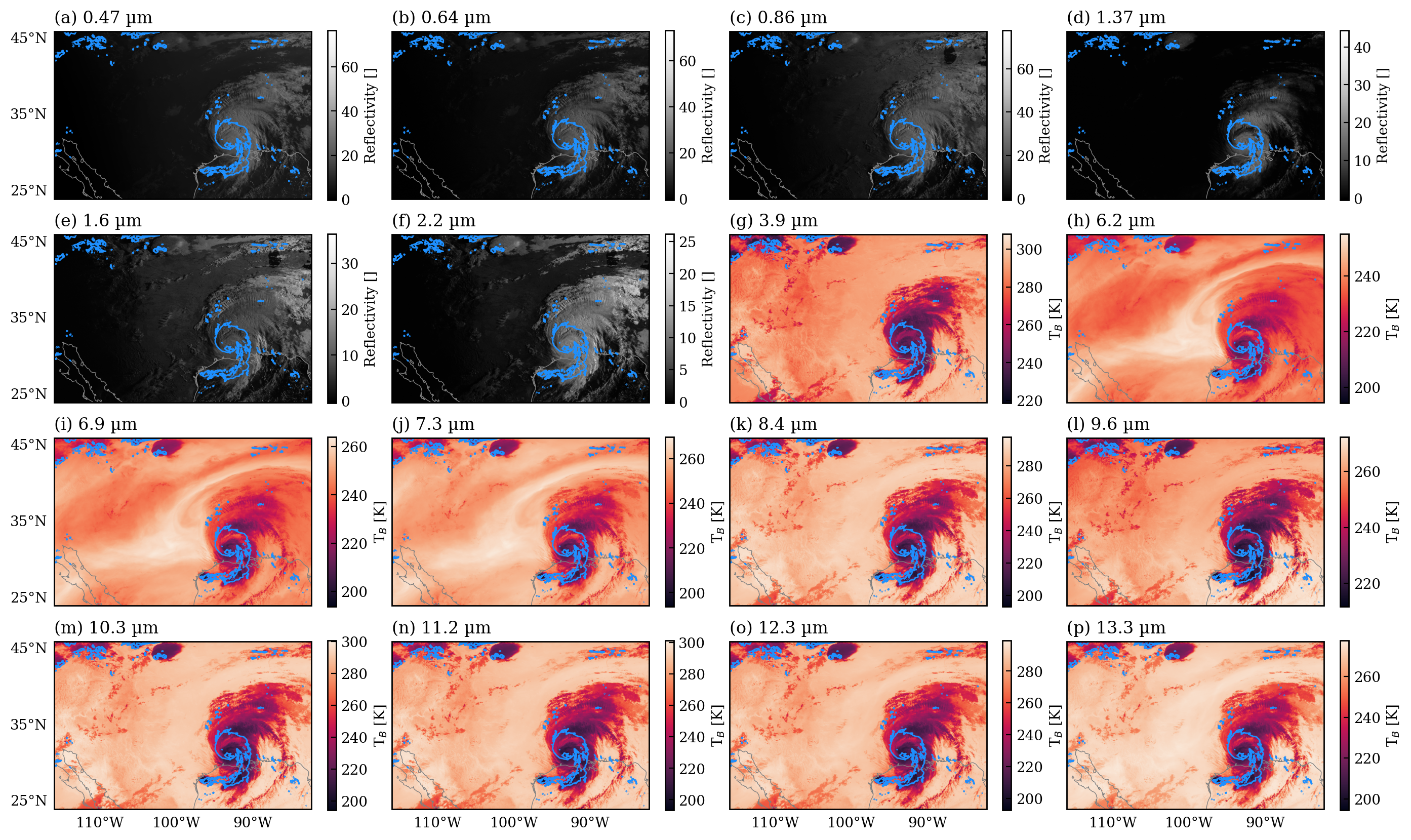}
	\caption{
		The 16 channels of the GOES ABI observing Hurricane Laura during
		landfall on August 27, 2020, at 08:41 UTC. The first six channels
		primarily measure reflected solar radiation. The remaining channels
		measure infrared radiation emitted from the atmosphere. Observations
		from geostationary sensors have the advantage of providing close-to
		continuous coverage but are sensitive primarily to cloud-top properties
		and thus only provide limited information on precipitation close to the
		surface. The blue contour encloses areas where surface precipitation
		exceeds \SI{1}{\milli \meter \per \hour}.
	}
	\label{fig:observations_geo}
\end{figure}

Visible (Vis) and infrared (IR) sensors benefit from the shorter wavelengths of
the radiation they measure, enabling much higher spatial resolution and
deployment on geostationary platforms. From this vantage point, they provide
near-continuous coverage of the underlying hemisphere. Consequently,
geostationary observations serve as a critical complement to PMW observations
for real-time and continuous precipitation monitoring.

Figure 3 shows observations from the 16 spectral channels of the ABI sensor
aboard GOES-16 during the landfall of Hurricane Laura. The native spatial
resolution of ABI channels ranges from 500 meters to 2 kilometers at the
sub-satellite point on the equator. Although this resolution degrades over the
Continental United States due to increasing viewing angles, it remains higher
than that of PMW sensors. The first six channels are visible and near-infrared
bands that measure reflected solar radiation and are only available during
daylight hours. The remaining ten channels operate in the thermal IR and provide
data continuously, both day and night. Among the thermal IR bands, the main
distinguishing feature is their sensitivity to atmospheric water vapor. For
instance, channels centered at 6.2, 6.9, and $\SI{7.3}{\micro \meter}$ are more
sensitive to upper-tropospheric moisture. This sensitivity to water vapor
provides contextual information on the moisture content of the air but reduces
the penetration depth of the observations.

A particular challenge related to incorporating geostationary observations into
global precipitation retrievals is that the channels availability changes
between platforms and thus the geographical coverage regions. SatRain integrates
data from CONUS, the Korean peninsula, and Austria, which are covered by
different geostationary platforms. While observations over CONUS are derived
from the GOES 16 \citep{Goodman2019Goes} platform, observations over Korea are
derived from the Himawari 8 and 9 platforms \citep{Bessho2016Himawari}, and
observations over Austria from the Meteosat 10 platform
\citep{Schmetz2002Introduction}. Table~\ref{tab:geo_channels} lists the central
wavelengths and corresponding bandwidths for the channels included from the ABI
sensor on GOES 16, the Advanced Himaware Imager (AHI,
\citeauthor{Da2015Preliminary}, \citeyear{Da2015Preliminary}), and the SEVIRI
\citep{Aminou2002Msg} sensor on Meteosat 10. The observations from the ABI, AHI,
and SEVIRI sensors were downloaded from \cite{NOAA_GOES_AWS},
\cite{NOAA_HIMAWARI_AWS}, and \cite{EUMETSAT_SEVIRI}, respectively..

\begin{table}[htbp]
  \centering
    \begin{tabular}{lccc}
      \toprule
      Channel & ABI & AHI & SEVIRI \\
      \midrule
      1 & 0.47 & 0.455 & 0.75  \\
      2 & 0.64 & 0.51  & 0.635 \\
      3 & 0.86 & 0.645 & 0.81  \\
      4 & 1.38 & 0.86  & 1.64  \\
      5 & 1.61 & 1.61  & 3.92  \\
      6 & 2.26 & 2.26  & 6.25  \\
      7 & 3.9  & 3.85  & 7.35  \\
      8 & 6.15 & 6.25  & 8.7   \\
      9 & 7.0  & 6.95  & 9.66  \\
      10 & 7.4  & 7.35  & 10.8 \\
      11 & 8.5  & 8.6   & 12.0 \\
      12 & 9.7  & 9.63  & 13.4 \\
      13 & 10.3 & 10.45 &    \\
      14 & 11.2 & 11.20 &    \\
      15 & 12.3 & 12.35 &    \\
      16 & 13.3 & 13.3  &    \\
      \bottomrule
	\end{tabular}
	\caption{Channel central wavelengths ABI sensor on GOES-16, the AHI sensor on Himawari-8/9, and the SEVIRI sensor on Meteosat-10.}
	\label{tab:geo_channels}
\end{table}

Latest-generation geostationary platforms provide observations at temporal
resolution of at least 10 minutes allowing them to closely track the evolution of
precipitation systems. In order to allow users to explore the temporal
information content in time-resolved geostationary observations, the SatRain
dataset includes observations from multiple time steps around the overpass of
the PMW sensor.

In addition to the multi-channel Vis and IR observations from the latest
generation of geostationary sensors, the SatRain dataset also integrates
$\SI{0.036}{\degree}$ gridded thermal IR observations from the \SI{11}{\micro \meter}
infrared window sourced from the Climate Prediction Center (CPC) global gridded
geostationary IR dataset \citep{NCEP_CPC_L3_IR}. Since these observations are
available almost continuously from 1998, they play an important role for
generating long-term precipitation records and are included as an independent
input data source in the SatRain dataset.

\subsubsection{Ancillary Data}

\begin{figure}[htbp] 
	\centering
	\includegraphics[width=1.0\textwidth]{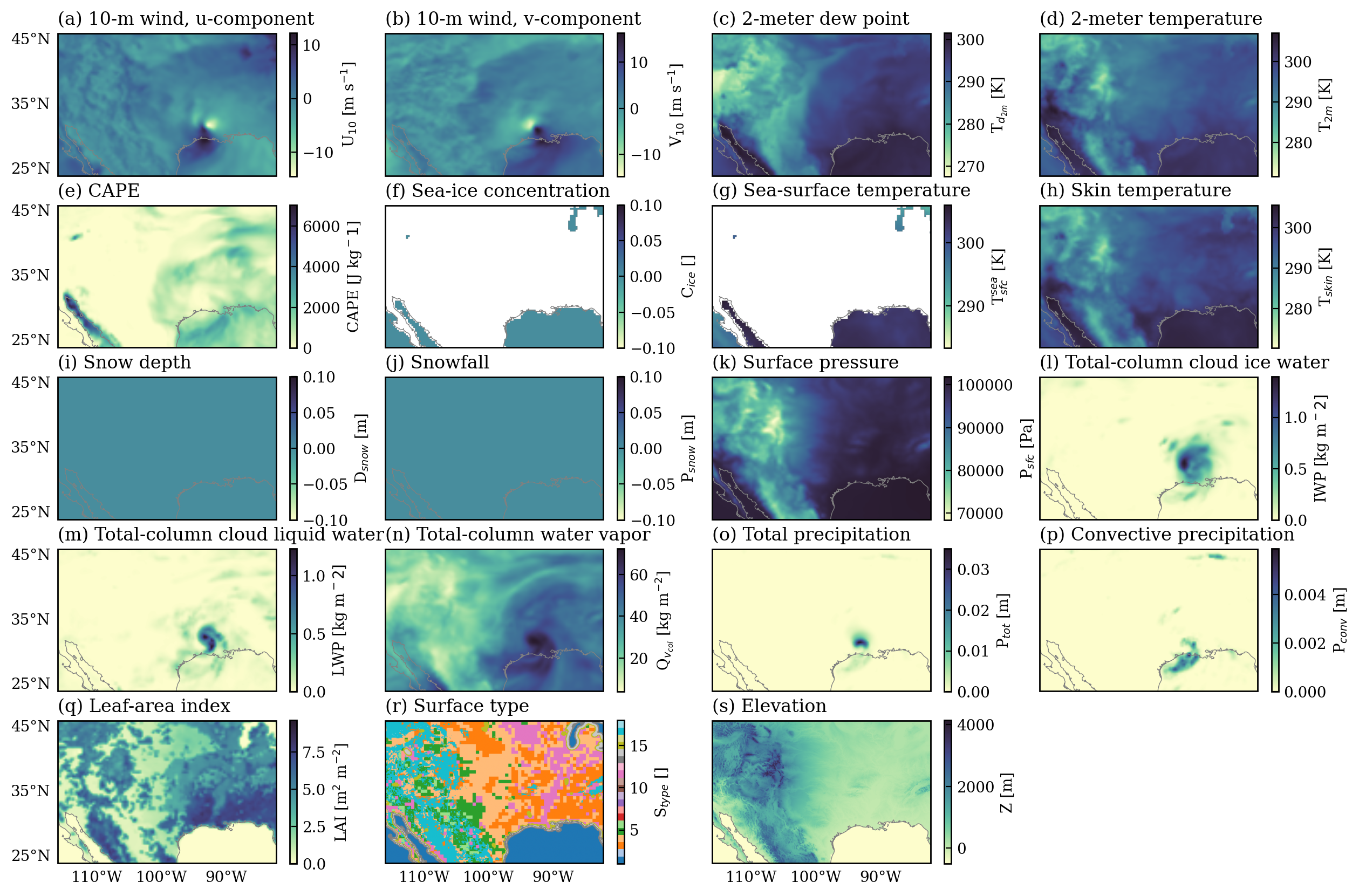}
	\caption{
		Ancillary variables provided by the SatRain dataset for the scene
		depicted in Fig. 1. Panels (a) to (q) show the dynamic ERA5 fields included in
		the ancillary data. Panel (r ) shows the 18-class surface classification from
		the GPROF algorithm and panel (s) the surface elevation from the NOAA Globe
		dataset.
	}
	\label{fig:ancillary_data}
\end{figure}

Because the relationship between satellite observations and surface
precipitation is often under-constrained, it is common to augment satellite
observations with complementary environmental information, so-called ancillary
data, to improve the accuracy of the precipitation estimates. Typical examples
include the surface type, atmospheric and surface temperatures, humidity, and
elevation. The SatRain dataset includes several static and dynamic ancillary
variables. Dynamic ancillary data describing the state of the atmosphere and the
surface are derived from the ERA5 \citep{Hersbach2020ERA5} dataset. In addition
to that, the ancillary data also contains an 18-class surface classification
that has been developed for the Goddard Profiling Algorithm (GPROF,
\citeauthor{GPM_GPROF_ATBD_V7}, \citeyear{GPM_GPROF_ATBD_V7}) precipitation
retrieval combing microwave-based surface-type information with snow- and
sea-ice coverage data from the Autosnow product \citep{NCEI2025SNOWMaps}. In
terms of static variables, SatRain provides the surface elevation sourced from
the NOAA Global Land One-kilometer Base Elevation (GLOBE) digital elevation
model \citep{Hastings1999GLOBE}. Figure~\ref{fig:ancillary_data} provides
showcases the ancillary variables included in the SatRain dataset for the
landfall of Hurricane Laura.

\subsubsection{Precipitation Reference}

\begin{figure}[htbp] 
	\centering
	\includegraphics[width=1.0\textwidth]{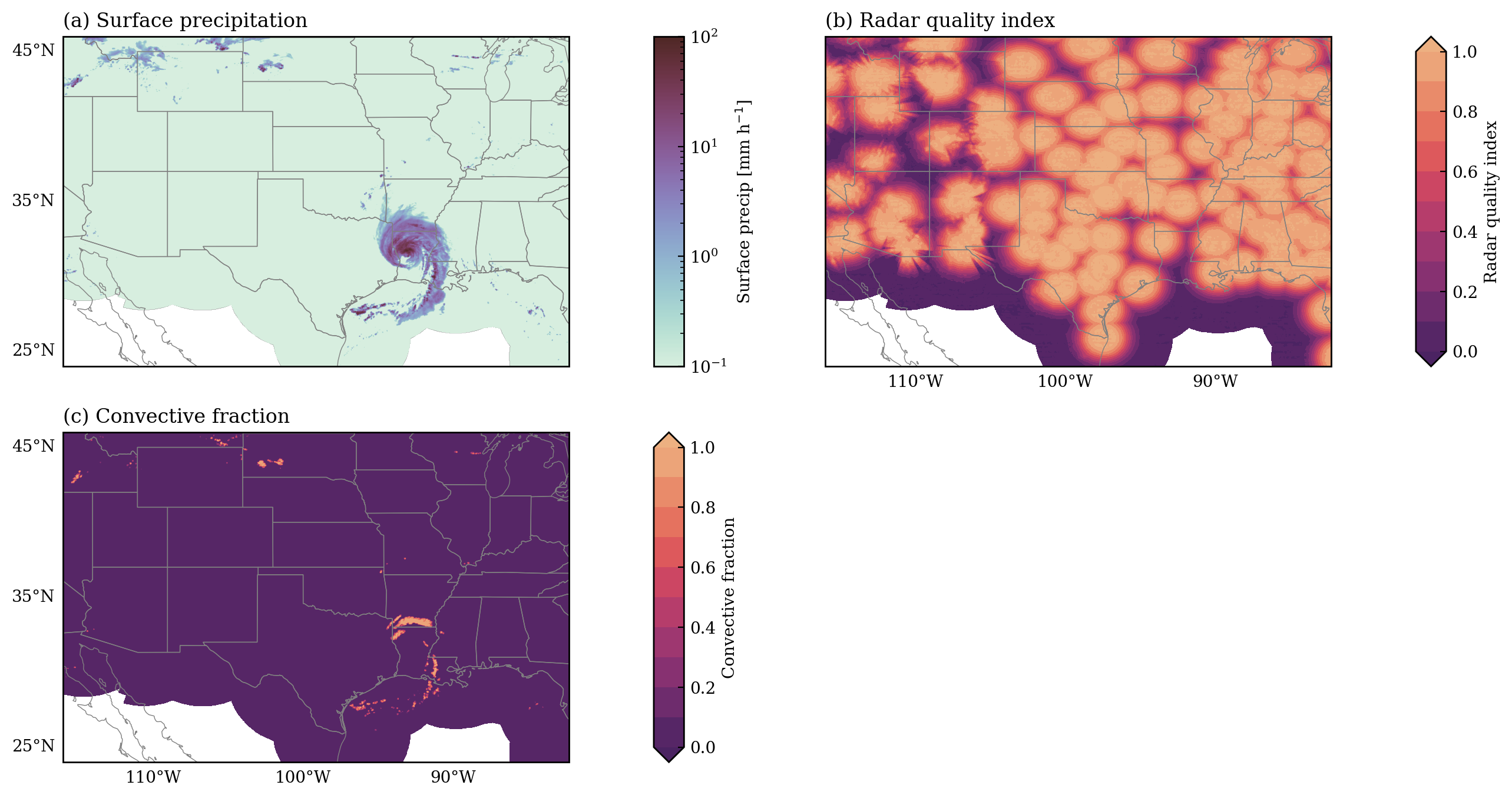}
	\caption{
		Reference precipitation estimates and auxiliary fields dataset during
		the landfall of Hurricane Laura. Panel (a) shows surface precipitation
		estimates from NOAA’s gauge-corrected Multi-Radar Multi-Sensor product.
		Panel (b) displays the radar quality index quantifying the reliability
		of the precipitation estimates. Panel (c) presents the convective
		fraction field, illustrating the hydrometeor classification data
		included in the SatRain dataset.
	}
	\label{fig:observations_geo}
\end{figure}

The precipitation reference used in the SatRain dataset are derived from
gauge-corrected ground-based precipitation radar measurements from NOAAs
gauge-corrected Multi-Radar Multi-Sensor (MRMS, \citeauthor{Smith2016MRMS},
\citeyear{Smith2016MRMS}) product. MRMS is based on radar observations from
Nexrad, the most extensive network of precipitation radars in the world
comprising around 160 polarimetric S-band radars. The radar-derived estimates of
liquid precipitation are corrected using hourly gauge-correction factors thus
enforcing consistency between instantaneous estimates and direct measurements of
hourly accumulations from gauge stations. While a certain level of residual
uncertainty in these estimates cannot be eliminated, they are generally
considered to be the best currently available estimates of surface precipitation
with near-complete coverage over CONUS. In addition to reference surface
precipitation rates, the SatRain dataset contains a radar quality index and the
gauge correction factor, allowing the user to customize the quality requirements
for the radar estimates used during both training and evaluation. Furthermore,
the dataset contains precipitation-type masks identifying convective and
stratiform rain, snow, and hail as provided by the MRMS data. Examples of these
fields are shown in Fig. 5.

\subsubsection{Independent Precipitation Reference}

\begin{figure}[htbp] 
	\centering
	\includegraphics[width=1.0\textwidth]{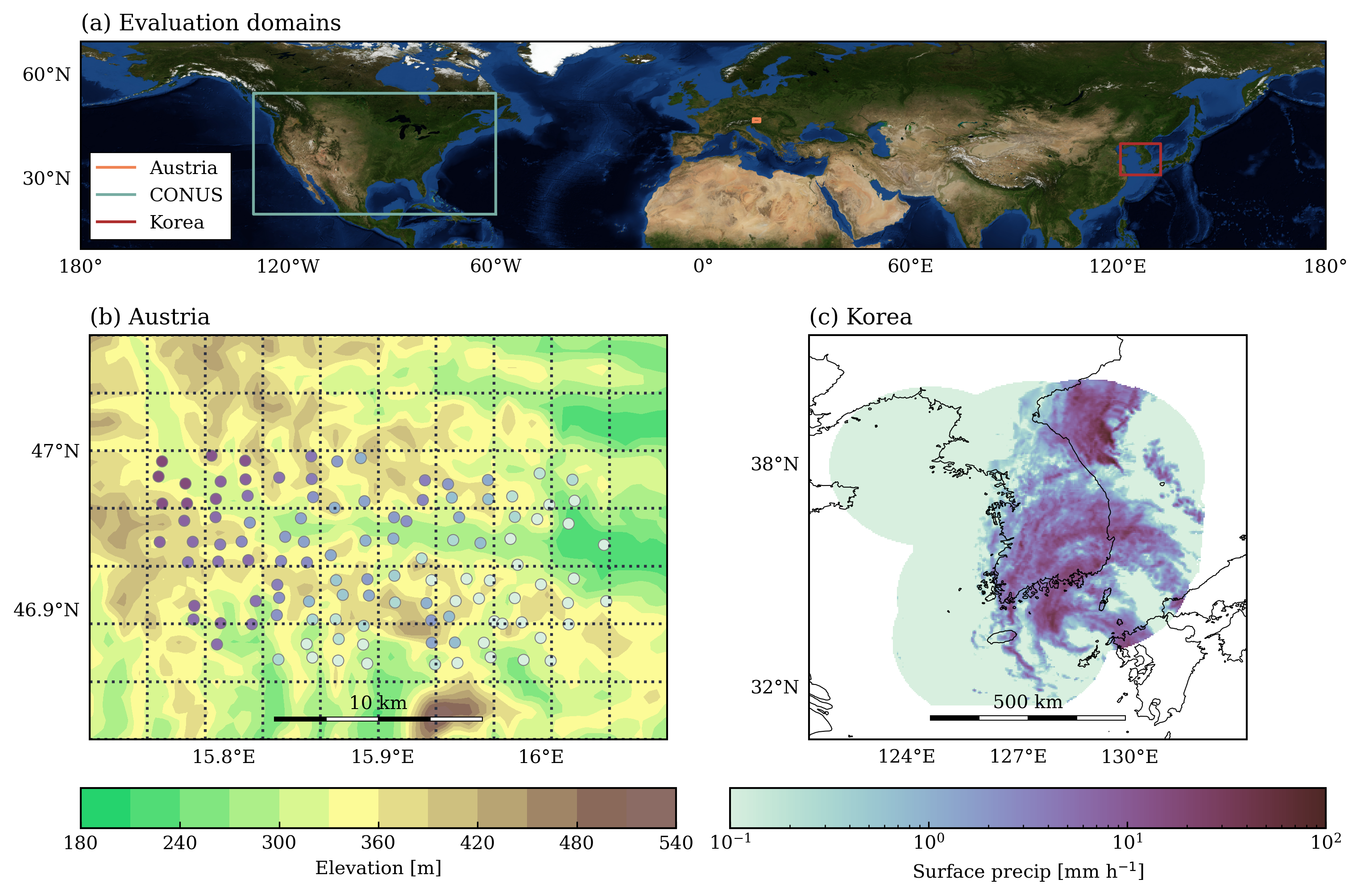}
	\caption{
		Independent test data included in the SatRain dataset. Panel (a) shows the
		spatial coverage of the three temporally and spatially independent test
		datasets. Panel (b) shows the location of the WegenerNet gauge stations used as
		reference measurements in the Austria domain and the grid they are aggregated
		to. Panel (c) shows an example in the Korea domain, with radar and rain-gauge
		data obtained from the Korea Meteorological Administration.
	}
	\label{fig:observations_geo}
\end{figure}

The SatRain training data is derived from four years (2018–2021) of ground-based
radar measurements over CONUS. While this region encompasses a relatively wide
range of climate zones, it does not fully capture the diversity of precipitation
systems observed globally. To mitigate this limitation, the testing data
includes a full year withheld from the training period, as well as two
additional test sets derived from distinct geographical regions and independent
measurement systems. These complementary datasets help reduce the risk of
overfitting to the weather patterns characteristic of the training domain.

The first geographically independent test dataset consists of gauge-corrected,
ground-based radar rainfall estimates over South Korea, generated using radar
compositing methods optimized for the Korean domain
\citep{Kwon2015RadarComparison}. The Korea-specific merging technique uses
radial basis function interpolation to combine radar and rain gauge observations
resulting in a high-resolution rainfall product in both space and time
\citep{Ryu2025RBFMerging}. Although the measurement approach is similar to that
used in the MRMS reference data over CONUS, the processing of raw data into
precipitation estimates differs. The dataset also represents a distinct climate
regime from the training data, offering a valuable test of the precipitation
retrieval model’s ability to generalize beyond the conditions typically
encountered over CONUS. Studies such as \cite{Sohn2013WarmRain} showed that
cloud characteristics affecting both PMW radiometers and IR observations differ
substantially between Korea and CONUS and can cause significant biases in
algorithms applied to these two regions. The independent testing data from Korea
thus provides an opportunity to assess whether retrieval improvements are
achieved at the cost of global generalizability.

The second independent evaluation dataset is derived from gauge measurements
from the WegenerNet \citep{Fuchsberger2021WegenerNet} gauge network around the
Feldbach region in Austria. While IPWG focuses primarily on gauge-corrected
radar data for validation, WegenerNet is unique in that its gauge density is
sufficiently high that the addition of radar data would not modify any of the
gauge accumulations. In addition, it offers validation over a mountainous regime
that radar/gauge networks still struggle with. For the comparison against
satellite-based precipitation estimates, the half-hourly accumulations from the
ground stations were converted to precipitation rates and aggregated to the
$\SI{0.036}{\degree}$ grid using binning. As can be seen in Panel (b) of Fig. 6,
the gauge density is sufficient to cover 24 grid cells with at least two gauges
per cell.

\subsection{Dataset Generation}

The SatRain dataset is generated by extracting observations from all available
overpasses of the GMI and ATMS sensors over the target domain and adding the
corresponding geostationary observations, ancillary data, and ground-based
reference precipitation estimates. The resulting collocation scenes are then
used to extract fixed-size training scenes to produce AI-ready training and
validation sets suitable for training ML precipitation retrievals.

The SatRain dataset is partitioned into training, validation, and testing
subsets that are designed to always be either temporally or spatially
independent. Training and validation data are extracted from the years 2019
through 2021, with the first five days of each month allocated to the validation
set and the remaining days to the training set. For testing, the CONUS subset
uses data from 2022. The independent test set over Austria is based on
observations from 2021 and 2022, while the Korea test set covers the period from
October 2022 through October 2023.

\subsubsection{Generation of Collocation Scenes}

\begin{figure}[htbp]
	\centering
	\includegraphics[width=1.0\textwidth]{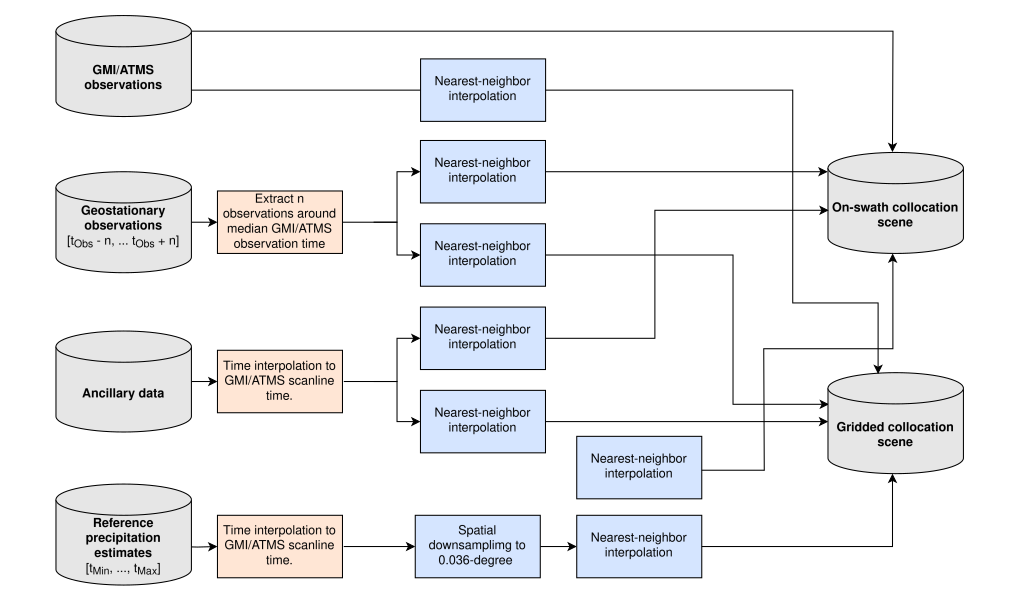}
	\caption{
		Flow diagram illustrating the data flow for collocating the satellite
		observations and ground-based reference precipitation estimates for the SatRain
		dataset.
	}
	\label{fig:data_flow}
\end{figure}

The first step of the creation of the SatRain dataset consists of the extraction
of collocation scenes for every overpass of the PMW base sensor, i.e., GMI or
ATMS, over the targeted domain (CONUS, Austria, or Korea). The resulting
collocation scene contains all retrieval input data, i.e. satellite observations
and ancillary data, combined with the coincident reference precipitation on a
shared spatial grid. Two type of collocation scenes are extracted for every
overpass of the base sensors: An on-swath scene, which uses the native sampling
of the PMW observations, and a gridded scene, which contains all data regridded
to a regular latitude-longitude grid with a resolution of $\SI{0.036}{\degree}$.

The collocation process, as illustrated in Figure~\ref{fig:data_flow}, starts
out with the PMW observation from an overpass of GMI or ATMS over the domain
containing the reference data. Corresponding ground-based reference data and
ancillary data are extracted to cover the time range of the overpass and
interpolated to the observation time of each scan-line of the PMW observations.
The MRMS data, which have a native resolution of $\SI{0.01}{\degree}$, are reduced in
resolution to match $\SI{0.036}{\degree}$ grid by smoothing using a Gaussian filter with
a full width at half maximum of $\SI{0.036}{\degree}$ and interpolated linearly to the
target grid. The mapping of the reference precipitation estimates to the
on-swath geometry is performed by nearest-neighbor interpolation. Similarly, the
PMW observations are mapped to the regular latitude-longitude grid by nearest
neighbor interpolation.

The global gridded IR geostationary observations are extracted at a temporal
resolution of $\SI{30}{\minute}$ over a time range of $\SI{8}{\hour}$ centered on the
median overpass time. Multi-channel Vis and IR observations from the ABI, AHI,
and SEVIRI sensors are extracted over a time window of 1 hour and a temporal
resolution of 10 minutes. The geostationary observations are mapped to the
gridded and on-swath geometries using nearest-neighbor interpolation.

\subsubsection{Training Scene Extraction}

To generate AI-ready training and validation data from the collocation dataset,
fixed-size training scenes are extracted from the previously created collocation
dataset, separately for the gridded and on-swath subsets. The scenes are
extracted randomly, allowing an overlap of up to 50\% between neighboring
scenes, requiring 75\% of the pixels to contain valid observations and reference
data. The scene size for the gridded data is 256 pixel $\times$ 256 pixel and 64
pixel $\times$ 64 pixel for the on-swath data. The smaller size of $64 \times
	64$ used for the ATMS on-swath data is due to lower across-track sampling rate
of ATMS, which records only 96 pixel per scans as opposed to 221 for GMI.

\begin{figure}[htbp] 
	\centering
	\includegraphics[width=1.0\textwidth]{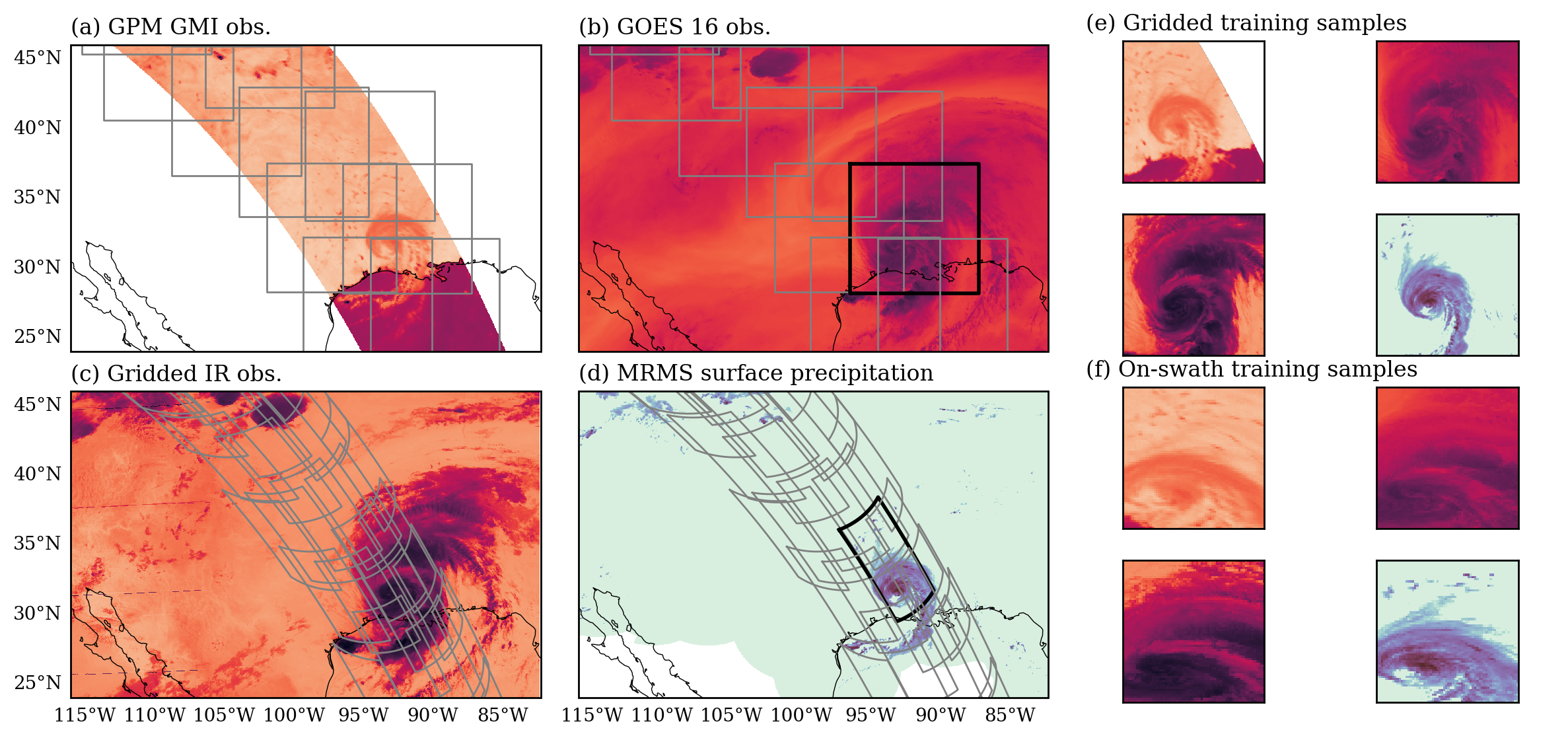}
	\caption{
		Fixed-size training scenes extracted from the SatRain collocation
		scenes. Panels (a), (b), and (c) show selected channels from the passive
		microwave (PMW), visible, and infrared observations that make up the
		input data of the SatRain dataset. Panel (d) shows the
		ground-radar-based precipitation reference used as training targets.
		Grey lines mark the outlines of the training samples extracted from this
		collocation scene for the gridded and on-swath observations. Black lines
		mark the sample training scenes displayed in Panel (e) and (f).
	}
	\label{fig:example_scenes}
\end{figure}

Figure~\ref{fig:example_scenes} shows an example collocation scene from a GMI
overpass of Hurricane Laura at 12:41 UTC on August 28, 2020. The grey boxes mark
the randomly extracted fixed-size training scenes in both the gridded and
on-swath geometries. Due to the irregular sampling pattern of PMW sensors, the
on-swath scenes appear distorted when displayed in the equirectangular (Plate
Carrée) projection used for the figures.

\section{Data Records}

The SatRain dataset is openly available for manual download from
\citep{SatRain2025}. Additionally, automated download functionality is provided
by the \texttt{satrain} \citep{SatRainPython2025} Python package.

\subsection{Dataset Organization}

The SatRain dataset is organized in a hierarchical folder structure that mirrors
its conceptual structure. At the top level, it is divided into two independent
subsets corresponding to collocation scenes from GMI overpasses and ATMS
observations, with the underlying PMW instrument referred to as the \textit{base sensor}.
For each base sensor, the data are partitioned into training, validation, and
testing splits in line with machine learning best practices. The validation
split contains samples from the first five days of each month, while the
training split includes the remainder. Although the validation set is generally
intended for monitoring model performance, users may also merge it with the
training data if desired.

To accommodate different computational and storage constraints, the training and
validation data are further subdivided into size-based subsets ('\texttt{xs}',
'\texttt{s}', '\texttt{m}', '\texttt{l}', and '\texttt{xl}'). This design
supports lightweight experimentation on smaller systems as well as large-scale
training on modern deep learning architectures. Files are not repeated across
the size-based subsets, instead each of the subsets is meant to include the
files in the preceeding, smaller subsets. Users aiming to use the \texttt{xl}
subset should thus combine files from all subsets.

\begin{figure}[hbpt!]
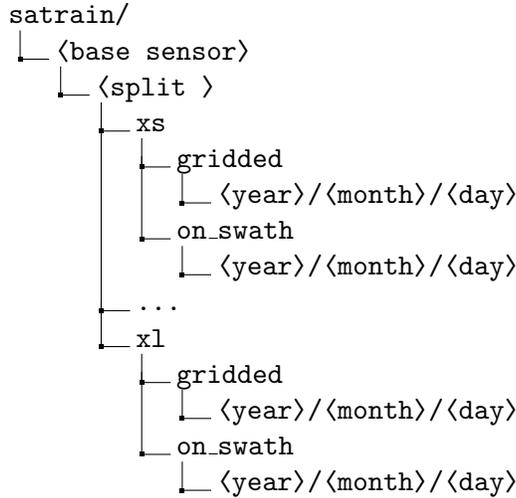
 

	\dirtree{%
		.1 satrain/.
		.2 \textlangle base sensor\textrangle.
		.3 \textlangle split \textrangle.
		.4 xs.
		.5 gridded.
		.6 \textlangle year\textrangle /\textlangle month\textrangle /\textlangle day\textrangle .
		.5 on\_swath.
		.6 \textlangle year\textrangle /\textlangle month\textrangle /\textlangle day\textrangle.
		.4 ....
		.4 xl.
		.5 gridded.
		.6 \textlangle year\textrangle /\textlangle month\textrangle /\textlangle day\textrangle.
		.5 on\_swath.
		.6 \textlangle year\textrangle /\textlangle month\textrangle /\textlangle day\textrangle.
	}

	\caption{
		Directory structure of the training and validation splits.
	}
	\label{fig:train_val_organization}
\end{figure}

Since all data is available in both gridded and on-swath geometries, the
directory is split once more into the data represented using the on-swath and
gridded spatial sampling. Finally, files are organized temporally into folders
by year, month, and day of the month. The resulting folder hierarchy for the
training and validation data is shown in Fig.~\ref{fig:train_val_organization}.

\begin{figure}[hbpt!]
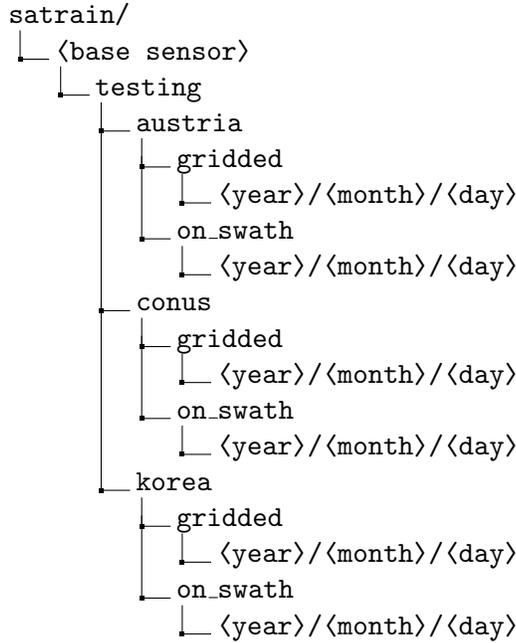
 

	\dirtree{%
		.1 satrain/.
		.2 \textlangle base sensor\textrangle.
		.3  testing .
		.4 austria.
		.5 gridded.
		.6 \textlangle year\textrangle /\textlangle month\textrangle /\textlangle day\textrangle .
		.5 on\_swath.
		.6 \textlangle year\textrangle /\textlangle month\textrangle /\textlangle day\textrangle.
		.4 conus.
		.5 gridded.
		.6 \textlangle year\textrangle /\textlangle month\textrangle /\textlangle day\textrangle .
		.5 on\_swath.
		.6 \textlangle year\textrangle /\textlangle month\textrangle /\textlangle day\textrangle.
		.4 korea.
		.5 gridded.
		.6 \textlangle year\textrangle /\textlangle month\textrangle /\textlangle day\textrangle .
		.5 on\_swath.
		.6 \textlangle year\textrangle /\textlangle month\textrangle /\textlangle day\textrangle.
	}

	\caption{
		Directory structure of the testing split.
	}
	\label{fig:testing_organization}
\end{figure}

The testing data are organized slightly differently from the training and
validation splits. The testing set is not subdivided into size-based subsets,
since all evaluations must be performed on the same data to ensure
comparability. Instead, the testing data are grouped by the underlying spatial
domain: CONUS, Korea, and Austria, which correspond to the available reference
data sources: MRMS over CONUS, ground-based radar data over Korea, and in-situ
measurements from the WegenerNet stations in Austria. The resulting folder
hierarchy is illustrated in Fig.~\ref{fig:testing_organization}.

A further distinction between the training and validation data on one hand and
the testing data on the other is that the collocation scenes in the testing set
are not divided into fixed-size patches. The testing data preserves the original
observation structure is to avoid sampling distortions, simplify comparisons
between gridded and on-swath retrievals, and enable direct evaluation of
existing precipitation products on the test data.

\subsection{Data Files}

The various input and target data for each training scene are stored in separate
NetCDF4 files. This modular organization allows users to download just the data
they intend to use, for example, only the small ('\texttt{s}') subset of the GMI
and target data to train and evaluate a retrieval using only passive microwave
observations. Each file is identified using an individual prefix
('\texttt{gmi\_}', '\texttt{atms\_}', '\texttt{geo\_}', '\texttt{geo\_ir}',
'\texttt{ancillary\_}', '\texttt{target\_}') following a time stamp in the format
\texttt{YYYYMMDDHHMMSS} containing the median observation time. Input and target
files corresponding to a specific training, validation, or testing sample can
thus be identified using this timestamp.

Due to the large size of the time-resolved geostationary observations, the
geostationary observations are split into files containing only the observations
closest to the reference precipitation estimates and files containing
observations from multiple observations times. The multi-timestep observations
are stored in separate files with the suffix '\texttt{\_t}', i.e.,
'\texttt{geo\_t\_\textlangle timestamp\textrangle.nc}' and '\texttt{geo\_ir\_t\_\textlangle timestamp\textrangle.nc}'.

\subsubsection{PMW Observations}

The PMW observations in the SatRain dataset consist of observations from the GMI
sensor and the ATMS sensor on the NOAA-20 satellite. They are stored in files
labeled '\texttt{gmi\_\textlangle timestamp\textrangle.nc}' for the subset using
GMI as base sensor and, correspondingly, '\texttt{atms\_\textlangle
  timestamp\textrangle.nc}' for the subset using ATMS as base sensor. Each file
contains the brightness temperatures in Kelvin for each channel
(\texttt{observations}), the corresponding earth-incidence angles in degree
(\texttt{earth\_incidence\_angle}), and the observation time
(\texttt{scan\_time}) corresponding to each scan line. The channels included for
the GMI and ATMS sensors are listed in table~\ref{tab:pmw_channels}.

\subsection{Geostationary Vis and IR Observations}

The multi-channel, geostationary Vis and IR observations from the GOES,
Himawari, and Meteosat platforms are stored in files named
'\texttt{geo\_\textlangle timestamp\textrangle.nc}'. These files contain the
observations closest in time to the measurement time of the reference
precipitation estimates in the variable \texttt{observations}. The visible
channels are stored as reflectances while thermal IR channels are stored using
brightness temperatures. The geostationary data currently does not include
viewing angles so the user will have to add them manually. The multi-timestep
files (\texttt{geo\_t\_\textlangle timestamp \textrangle.nc}') contain
observations from four 10-minute time steps prior to the collocation median and
three 10-minute time steps after the collocation median time.

Since GOES observations are only available over CONUS, the geostationary
observations for the test data from the Korea and Austria domains are derived
from AHI sensor onboard Himawari 8 and 9 and the SEVIRI sensor onboard Meteosat
10. Table~\ref{tab:geo_channels} lists the central wavelengths of the
channels of each of the included sensors. Since the sensors have different
channels, the observations differ in their spectral coverage and users will need
to account for that in their algorithm design.

\subsection{Geostationary IR Observations}

The single-channel gridded IR observations from the global, merged CPC dataset
\citep{NCEP_CPC_L3_IR} are stored in files named '\texttt{geo\_ir\_\textlangle
	timestamp\textrangle.nc}' and contain IR window-channel observations from
wavelengths around $\SI{11}{\micro \meter}$. The observed brightness
temperatures in $\si{K}$ are stored in the variable \texttt{observations}. The
single-timestep files contain the GEO IR observations closest to the
measurement time of the reference precipitation estimates. The
temporally-resolved GEO IR observations contain 16 half-hourly observations
centered on the median observation time and are stored in files named
'\texttt{geo\_ir\_t\_\textlangle timestamp\textrangle.nc}'.

\subsection{Ancillary Data}

The ancillary data is stored in separate files named
'\texttt{ancillary\_\textlangle timestamp \textrangle.nc}'. Each file contains
the ancillary variables listed in table~\ref{tab:ancillary}.

\begin{table}[htbp]
	\centering
	\begin{tabular}{lp{4cm}ll}
		\toprule
		\textbf{Variable name}                       & \textbf{Explanation}                                   & \textbf{Unit}                               & \textbf{Source} \\
		\midrule
		\texttt{ten\_meter\_wind\_u}                 & Zonal wind at 10 m altitude                            & $\si{\meter \per \second}$                  & ERA5            \\
		\texttt{ten\_meter\_wind\_v}                 & Meridional wind at 10 m altitude                       & $\si{\meter \per \second}$                  & ERA5            \\
		\texttt{two\_meter\_dew\_point}              & Dew-point temperature at 2 m altitude                  & $\si{\kelvin}$                              & ERA5            \\
		\texttt{two\_meter\_temperature}             & Near-surface temperature                               & $\si{\kelvin}$                              & ERA5            \\
		\texttt{cape}                                & Convective available potential energy                  & $\si{\joule \per \kilo \gram}$              & ERA5            \\
		\texttt{sea\_ice\_concentration}             & Fractional sea-ice coverage                            & --                                          & ERA5            \\
		\texttt{sea\_surface\_temperature}           & Sea surface temperature                                & $\si{\kelvin}$                              & ERA5            \\
		\texttt{skin\_temperature}                   & Surface skin temperature                               & $\si{\kelvin}$                              & ERA5            \\
		\texttt{snow\_depth}                         & Snow depth                                             & $\si{\kilo \gram \per \meter \squared}$     & ERA5            \\
		\texttt{snow\_fall}                          & Snowfall rate                                          & $\si{\meter \per \hour}$                    & ERA5            \\
		\texttt{surface\_pressure}                   & Surface pressure                                       & $\si{\hecto \pascal}$                       & ERA5            \\
		\texttt{total\_column\_cloud\_ice\_water}    & Vertically integrated mass of cloud ice                & $\si{\kilo \gram \per \meter}$              & ERA5            \\
		\texttt{total\_column\_cloud\_liquid\_water} & Vertically integrated mass of liquid cloud droplets    & $\si{\kilo \gram \per \meter \squared}$     & ERA5            \\
		\texttt{total\_column\_water\_vapor}         & Vertically integrated mass of water vapor              & $\si{\kilo \gram \per \meter \squared}$     & ERA5            \\
		\texttt{total\_precipitation}                & Total precipitation                                    & $\si{\meter \per \hour}$                    & ERA5            \\
		\texttt{convective\_precipitation}           & Convective precipitation                               & $\si{\meter \per \hour}$                    & ERA5            \\
		\texttt{leaf\_area\_index}                   & Half of the total green leaf area per unit ground area & $\si{\meter \squared \per \meter \squared}$ & ERA5            \\
		\texttt{surface\_type}                       & GPROF 18-class dynamic surface classification          & --                                          & GPROF V7        \\
		\texttt{elevation}                           & Surface elevation                                      & $\si{\meter}$                               & NOAA GLOBE      \\
		\bottomrule
	\end{tabular}
	\caption{List of the variables, their explanations, units, and data sources included in the ancillary data.}
	\label{tab:ancillary}
\end{table}

\subsection{Reference Data}

The reference data for every scene is stored in a file
'\texttt{target\_\textlangle timestamp\textrangle.nc}', where the timestamp
matches that of the input data. The reference data files contain the surface
precipitation in a variable called '\texttt{surface\_precip}'. Additionally, the
reference data files derived over CONUS contain several quality indicator
variables that can be used to filter the reference data samples used for
training and validation. The primary quality indicator is the radar quality
index (\texttt{radar\_quality\_index}), which provides an estimate of the
quality of the surface precipitation estimates based on the radar beam height,
beam blockage, and the height of the freezing level. Moreover, the files contain
the gauge correction factor (\texttt{gauge\_correction\_factor}) that was
applied to correct the radar-only precipitation estimates. Finally, the files
also contain the fraction of valid (\texttt{valid\_fraction}), snowing
(\texttt{snow\_fraction}), and hailing (\texttt{hail\_fraction})
$\SI{0.01}{\degree}$-resolution pixels within the downsampled $\SI{0.036}{\degree}$ grid box of
the SatRain dataset. Since the radar data over Korea and the station data over
Austria do not provide this additional information, these auxiliary fields are
not provided by the target data files from the Korea and Austria domains.

\section{Technical Validation}

To assess the technical consistency of the dataset, we trained three
precipitation retrieval models on the CONUS-based training data, each relying on
one of the primary input observation types: multi-channel geostationary
measurements from GOES ABI (GEO), single-channel infrared data from the CPCIR
dataset (GEO-IR), and passive microwave observations from GMI (GMI). These
models were then applied to estimate precipitation during the landfall of
Typhoon Khanun over Korea. All retrievals rely solely on satellite observations
and exclude the ancillary data included in SatRain. Each retrieval is
implemented using a UNet architecture built on EfficientNet-V2 with 10 million
parameters.

\begin{figure}[htbp] 
	\centering
	\includegraphics[width=1.0\textwidth]{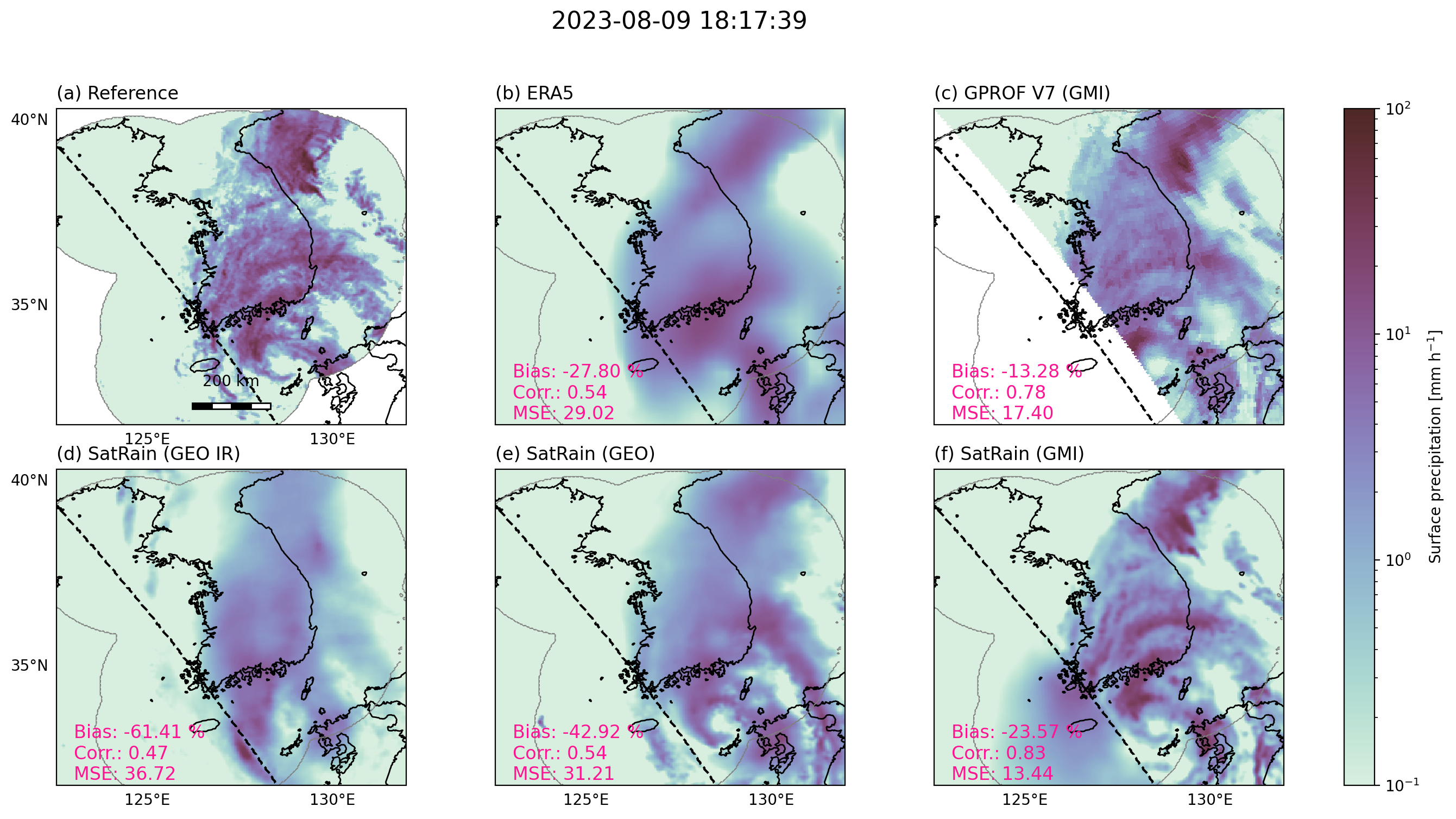}
	\caption{
		Precipitation retrievals trained using the CONUS-based SatRain
		training dataset compared to ground-based precipitation estimates and two
		baseline datasets during the landfall of Typhoon Khanun over the Korean
		peninsula. Panel (a) shows the ground-based reference precipitation
		estimates, panel (b) shows the precipitation field from the ERA5 reanalysis,
		panel (c) shows precipitation estimates from the GPROF precipitation
		retrieval applied to GMI observations. Panels (d), (e), and (f) show
		retrievals trained using the SatRain dataset over CONUS using multi-channel
		Vis/IR observation, single-channel GEO-IR observations, and GMI PMW
		observations, respectively. The dashed line marks the bounadary of the GMI
		swath.
	}
	\label{fig:case_study}
\end{figure}

The retrieved precipitation fields are compared to the reference ground-based
radar estimates and two baselines from the ERA-5 reanalysis dataset and the
GPROF precipitation retrieval in Fig.~\ref{fig:case_study}. The results
demonstrate that all retrievals capture the main precipitation structures of
Typhoon Khanun, but with clear differences in accuracy reflecting the
information content of their input observations \citep{Kidd2011GPM}. The
PMW-based retrieval performs best, reproducing much of the fine-scale structure
evident in the reference estimates. The multi-channel geostationary retrieval
captures the primary precipitation bands but misses finer details resolved by
PMW. In contrast, the single-channel IR retrieval shows the weakest performance,
with limited structural detail and correspondingly lower linear correlation and
higher mean-squared error.

Compared to the conventional baselines, the SatRain GMI retrieval shows visibly
better agreement with the reference precipitation fields than GPROF V7, a result
that is supported by scene-based accuracy metrics. The SatRain Vis and IR
retrieval achieves accuracy comparable to ERA5 but worse than GPROF V7, which is
expected since GPROF V7 leverages passive microwave observations that provide a
more direct link to precipitation. The SatRain Geo-IR retrieval performs worse
than both baseline for methods this specific case. However, this is consistent
with the very limited information content available from its input observations.

\begin{figure}[htbp] 
	\centering
	\includegraphics[width=1.0\textwidth]{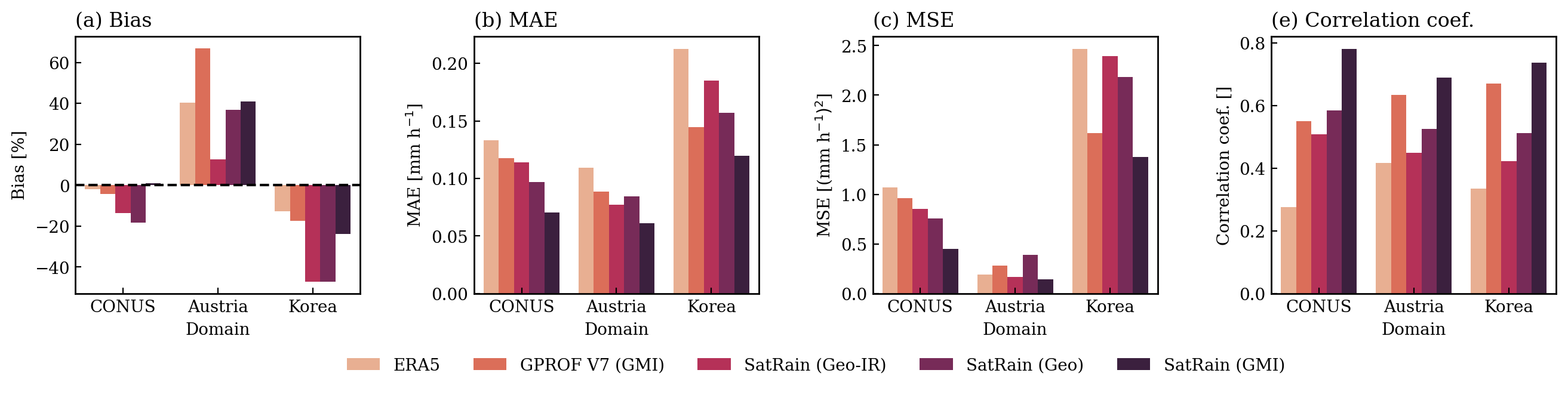}
	\caption{
		Evaluation of quantitative precipitation estimates from two state-of-the-art
		precipitation datasets (ERA5 and GPROF) and ML retrievals trained using the
		different satellite observations of the SatRain dataset. Each retrieval is
		evaluated for each of the three testing domains of the SatRain dataset. Panels
		(a) to (d) show the Bias, mean-absolute error (MAE), mean-squared error (MSE),
		and linear correlation coefficient, respectively.
	}
	\label{fig:sensor_comparison}
\end{figure}

The models trained on the CONUS dataset were further evaluated against three
independent test domains (CONUS, Austria, and Korea). The results, displayed in
Fig.~\ref{fig:sensor_comparison}, are consistent with the case study presented
above. The SatRain-based GMI retrieval consistently outperforms the GPROF V7
baseline, despite relying on the same observations. Retrievals based on
geostationary measurements (GEO and GEO-IR) achieve lower accuracy overall but
still surpass the ERA5 baseline. These improvements across multiple domains
confirm both the technical soundness and the practical value of SatRain.

While the SatRain retrievals demonstrate notable biases over Austria and Korea,
these are likely attributable to the exclusive use of CONUS data for training.
Such systematic errors align with previous studies documenting the sensitivity
of precipitation retrievals to regional precipitation characteristics
\citep{Sohn2013WarmRain}. Importantly, these biases are not unique to SatRain
but also affect operational products such as GPROF. Beyond these biases, the
relative performance of SatRain versus the baseline retrievals mirrors the
results obtained over CONUS. Despite differences in absolute error magnitudes
across domains, the ranking of retrieval methods remains largely consistent,
suggesting that SatRain-based evaluations generalize well across independent
regions, sensors, and time periods.

One exception occurs for the GEO retrieval over Austria, where MSE and MAE
values increase relative to other domains. This degradation likely reflects
differences between the SEVIRI channels available over Austria and the ABI
channels on which the model was trained. Although training was limited to
channels with the closest overlap and quantile remapping was applied to
harmonize the distributions, discrepancies in channel characteristics and
calibration still reduced performance. Accommodating such inter-sensor
differences remains an open challenge for machine learning–based precipitation
retrievals, and the SatRain dataset offers a platform to investigate these
issues further.

In addition to cross-sensor comparisons, we also used SatRain to evaluate the
performance of different machine learning approaches. Using GMI and ATMS
observations, we trained four retrieval models based on Random Forests, XGBoost,
a multi-layer perceptron (MLP), and a convolutional neural network (CNN). Their
performance, shown in Fig.~\ref{fig:sensor_comparison}, reveals a consistent
ranking across both sensors and all three domains: CNN-based models clearly
outperform the other techniques. Although the relative ordering of Random
Forests, XGBoost, and MLP varies by region, their performance differences remain
modest. These findings underscore SatRain’s utility as a robust and reliable
benchmark for assessing the skill of machine learning–based precipitation
retrievals.

\begin{figure}[htbp]
	\centering
	\includegraphics[width=1.0\textwidth]{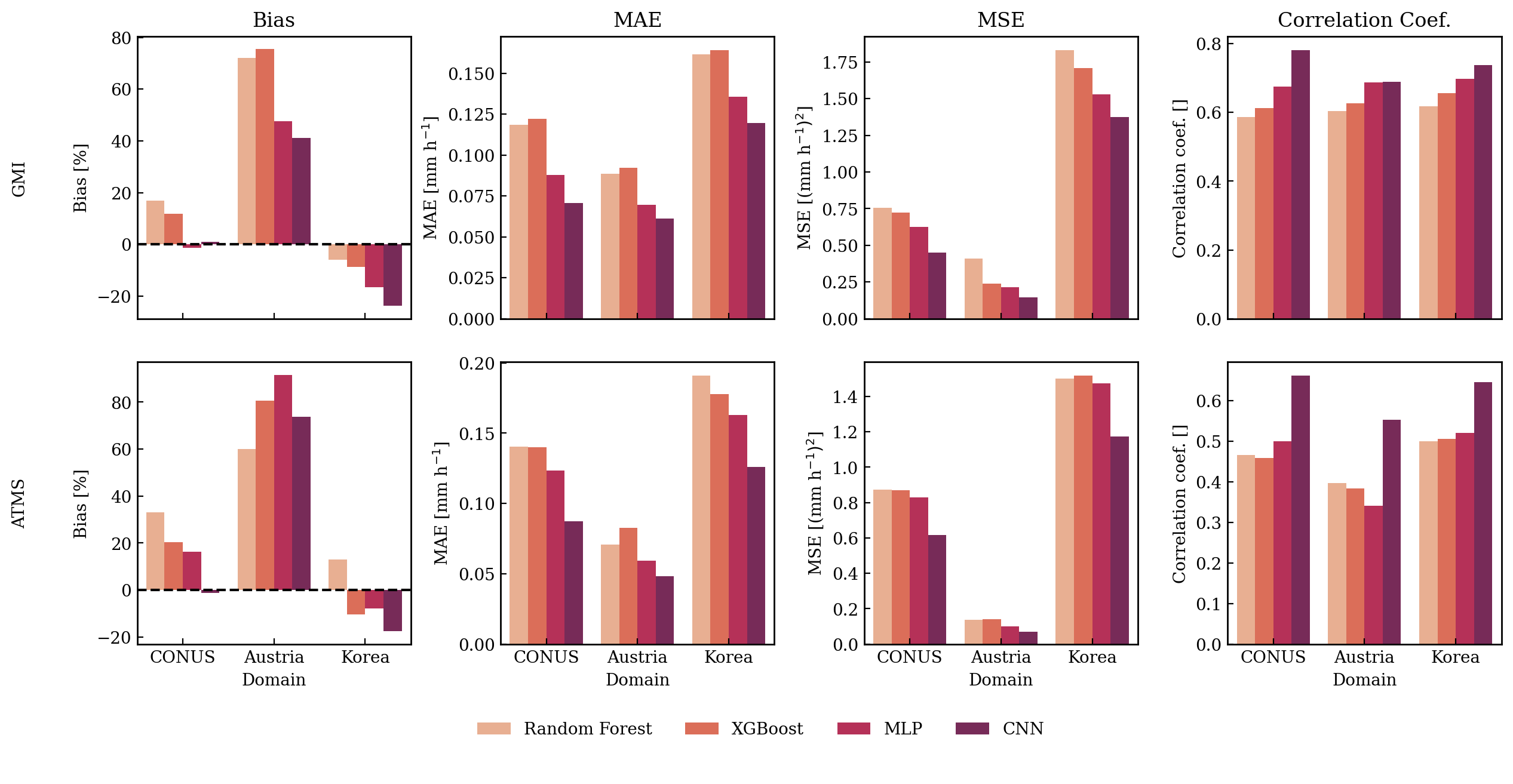}
	\caption{
		Like Fig. 8 but for four different machine-learning techniques trained on GMI (first row)
		and ATMS (second row) observations.
	}
	\label{fig:sensor_comparison}
\end{figure}

Taken together, these results underscore the value of SatRain as a benchmark
dataset. It offers a well-defined framework for training and evaluation that
enables direct comparison of different ML techniques. The independent test data
provide a crucial check, ensuring that performance gains generalize beyond the
training domain rather than reflecting overfitting to regional measurement
errors or precipitation characteristics. In addition, the ability to evaluate
models across sensor types and to combine observations from multiple sensors and
time periods makes SatRain a flexible foundation for developing advanced
ML-based retrieval methods.

\section{Usage Notes}

\begin{figure}[htbp]
	\centering
	\includegraphics[width=1.0\textwidth]{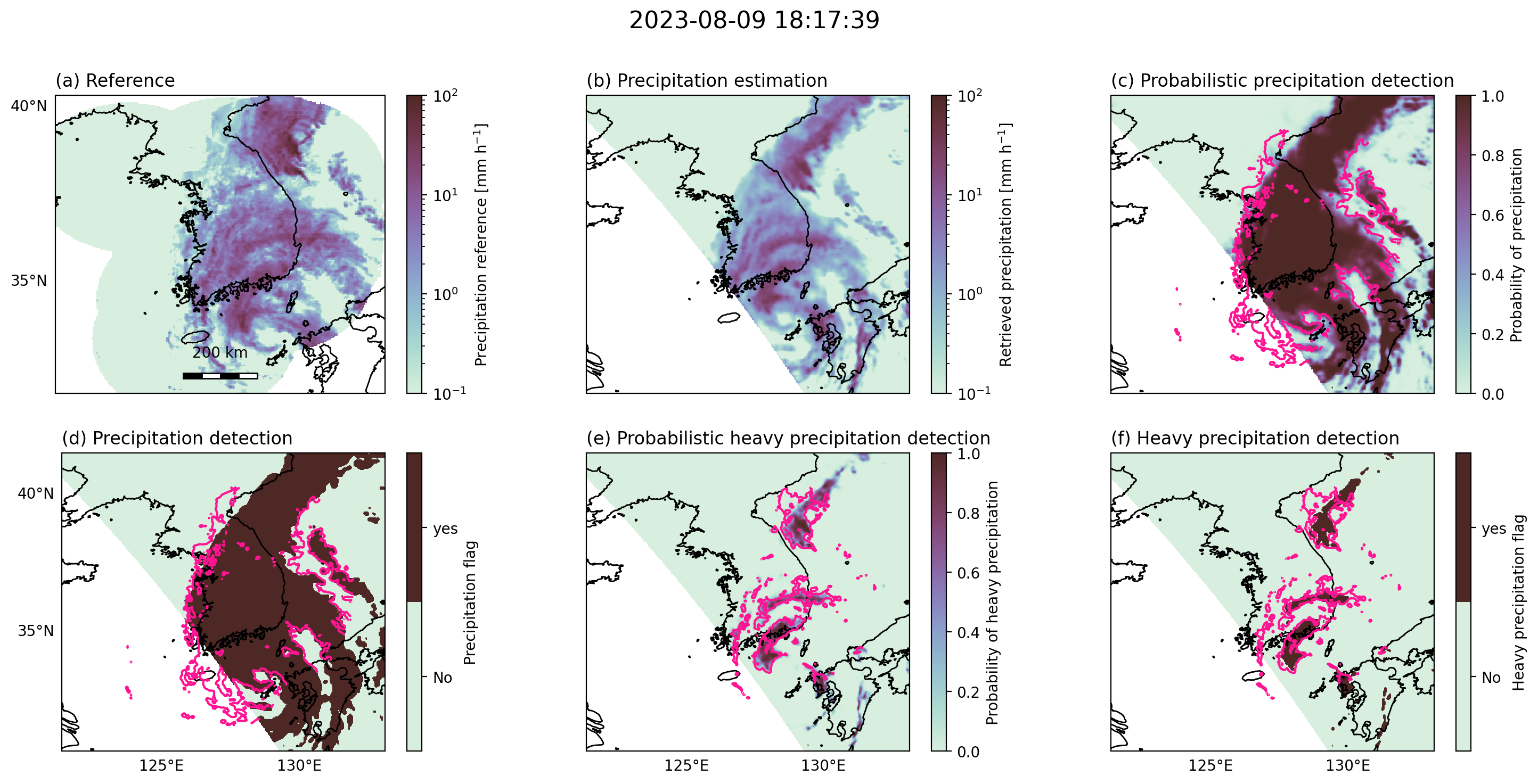}
	\caption{
		Example retrieval results for the five suggested precipitation estimation and
		detection tasks during landfall of Typhoon Khanun on August 9, 2023. Panel (b)
		shows quantitative precipitation estimates retrieved from GMI observations.
		Panel (c) shows the probability of precipitation for the probabilistic
		precipitation detection task. Panel (d) shows the precipitation flag for the
		deterministic precipitation detection task. Panel (e) and (f) show the
		corresponding probabilistic and deterministic results for the detection of
		heavy precipitation, defined as precipitation exceeding $\SI{10}{\milli \meter
				\per \hour}$.
	}
	\label{fig:tasks}
\end{figure}

Different scientific and societal applications may require precipitation
retrievals to emphasize specific characteristics, such as retrievals targeting
extreme precipitation or specific precipitation types such as snow or hail.
While no single benchmark can serve every possible use case, the SatRain dataset
defines five distinct evaluation tasks designed to test the ability of ML
algorithms to reproduce key aspects of precipitation events. These tasks
include: (1) precipitation rate estimation, (2) probabilistic detection of
precipitation, (3) deterministic detection of precipitation, (4) probabilistic
detection of heavy precipitation, and (5) deterministic detection of heavy
precipitation. We adopt thresholds of 0.1 mm/h and 10 mm/h to define
precipitation and heavy precipitation, respectively.
Figure~\ref{fig:tasks} illustrates example results for each of these
tasks using the GMI retrieval presented in the previous section.

\subsection{Evaluation Protocol}

To ensure fair comparison of precipitation retrievals trained on the SatRain
dataset, it is essential that models are evaluated using a consistent set of
criteria. We therefore propose a standardized evaluation protocol for
benchmarking ML retrievals on the SatRain dataset. Users may
choose to evaluate their models on all or a subset of the defined tasks. For
each task, the recommended accuracy metrics are listed in Table 4.

\begin{table}[htbp]
	\centering
	\caption{Precipitation estimation and detection tasks and metrics for evaluating
		precipitation retrieval methods on the SatRain dataset.}
	\begin{tabular}{>{\raggedright\arraybackslash}p{0.35\linewidth}
		>{\raggedright\arraybackslash}p{0.6\linewidth}} \toprule
		\textbf{Task}                               & \textbf{Metrics}                                   \\ \midrule Precipitation quantification &
		Relative bias, mean absolute error, mean squared error, symmetric mean
		absolute percentage error, linear correlation coefficient, effective
		resolution                                                                                       \\
		\addlinespace
		Precipitation detection                     & Probability of detection, false alarm rate, Heidke
		Skill Score                                                                                      \\
		\addlinespace
		Probabilistic precipitation detection       & Precision--recall curve                            \\
		\addlinespace
		Heavy precipitation detection               & Probability of detection, false alarm rate,
		Heidke Skill Score                                                                               \\
		\addlinespace
		Probabilistic heavy precipitation detection & Precision--recall curve
		\\ \bottomrule
	\end{tabular}
\end{table}

To ensure comparability of benchmark results across algorithms, evaluations
should be performed against the gridded reference data. This approach avoids
distortions in the evaluation statistics caused by the irregular spatial
sampling of on-swath observations. For cross-track scanners, for example, both
sampling density and spatial resolution decrease toward the swath edges. The
reduced sampling at the swath edges would thus underestimate the effect of the
reduced resolution compared to evaluation against the gridded data. To
facilitate reproducible comparisons, the gridded reference data files include the scan
and pixel coordinates of the nearest swath pixel of the corresponding on-swath file.
Using this information, the on-swath retrieval results can be mapped to the gridded
reference data easily and consistently.

A reference implementation of the proposed protocol is available through the
\texttt{satrain} package. By default, the evaluation compares all retrieval
outputs against the reference precipitation estimates on the 0.036° regular
latitude-longitude grid, ensuring results are independent of the retrieval’s
native coordinate system. For the CONUS domain, evaluations should be restricted
to regions with a radar-quality index of at least 0.5 and should include all
precipitation types. For the other two domains all reference estimates should be
used for the evaluation.

\subsection{The \texttt{satrain} package}

To facilitate community access and encourage widespread adoption, we developed
the \texttt{satrain} Python package. This package streamlines dataset download
and management, allowing users to begin working with SatRain without the burden
of manual data handling. Comprehensive documentation, including installation
instructions, usage examples, and tutorials, is available at
\url{satrain.readthedocs.org}.

In addition to data access, the package implements SatRain’s standardized
evaluation protocol. This enables users to benchmark models trained directly on
SatRain and to evaluate independently developed retrievals using the same
criteria. By providing a simple interface function, users can leverage the
package to automatically conduct evaluations across all testing domains, tasks,
and metrics. The \texttt{satrain} package thus offers a fast, consistent, and
reproducible framework for evaluating precipitation retrieval algorithms.

\subsection{Limitations}

The SatRain dataset is constructed from high-quality input datasets using
state-of-the-art techniques designed to reduce uncertainties in both the
satellite observations and the precipitation reference. Despite these efforts,
residual uncertainties and measurement errors remain in both components.

On the input side, clearly corrupted satellite imagery is flagged and removed
from the satellite-observation data used to construct the SatRain dataset.
However, more subtle issues such as undetected artifacts or gradual changes in
sensor characteristics may persist and affect the data. These represent
practical challenges that any precipitation retrieval must contend with
warranting their inclusion in the SatRain dataset.

The precipitation reference data are also subject to significant uncertainties.
While gauge-corrected, ground-based radar composites are widely regarded as the
most reliable spatially continuous precipitation estimates currently available,
they are not error-free. Beam overshooting, uncertainties in microphysical
processes, and the particular difficulty of estimating snowfall introduce
systematic biases. Snowfall poses a notable challenge: most gauges do not
accurately measure snow, and MRMS does not apply gauge correction to snowfall
estimates. Consequently, snowfall included in SatRain training and testing data
should be regarded as highly uncertain. For the WegenerNet test data, only
heated gauges are used under likely snowfall conditions, but limitations remain.

To reduce the impact of these uncertainties on retrieval evaluation, SatRain
includes independent test datasets from geographically distinct domains that
rely on entirely independent measurement systems. Performance gains that
transfer to these independent datasets are more likely to reflect genuine
improvements in retrieval capability, rather than overfitting to the same
reference data used in training.

It is also important to note that the primary purpose of SatRain is to serve as
a benchmark for evaluating and comparing retrieval algorithms, rather than as a
basis for developing globally accurate precipitation retrievals. The SatRain
dataset is limited to training data over the CONUS and is therefore not designed
for the development of global precipitation retrievals. Algorithms trained on
SatRain will learn to capture regional precipitation characteristics specific to
North America, which may result in substantial biases or retrieval errors when
applied to other parts of the world.

\subsection{Future Directions}

The SatRain dataset represents the first AI-ready benchmark for satellite-based
precipitation estimation and detection, marking an important step toward
facilitating the operational adoption of ML-based retrievals. It
brings together a comprehensive set of satellite observation types alongside
high-quality precipitation reference data derived according to best practices
agreed upon by the international community. By providing a common reference
point, SatRain enables systematic assessment of ML advances in
precipitation retrieval, improving the comparability, reproducibility, and
eventual uptake of retrieval techniques reported in the scientific literature.

Looking ahead, SatRain also opens opportunities for advancing precipitation
retrieval techniques beyond the currently dominant single-sensor approaches. Its
multi-sensor, multi-timestep design provides a strong foundation for developing
next-generation algorithms that fuse observations across sensors and time steps
to better capture precipitation structure and evolution. Furthermore, the
dataset is ideally suited for tackling broader challenges in global
precipitation retrieval, including the development of sensor-agnostic retrievals
and the mitigation of regional biases. By providing a common starting point to
address these challenges, SatRain can help guide the community toward more
robust, transferable, and operationally relevant precipitation retrieval
systems.

The SatRain project has been developed as a community effort to provide both an
accessible testbed for novel retrieval algorithms and a common reference for the
precipitation remote sensing community. The dataset and the software used to
generate it are openly available to encourage broad adoption and continued
development. From the outset, the dataset was designed with future extensions in
mind, and we invite contributions from the international community. Our hope is
that SatRain will serve as a foundation for stronger global collaboration aimed
at advancing satellite-based precipitation estimation and detection.

\section*{Data Availability}

\begin{itemize}
\item GPM L1C observations for GMI and ATMS were downloaded from \cite{Berg2022_GMI_L1C_R_V07} and \cite{Berg2022_ATMS_NOAA20_1C_V07}, respectively.
\item Observations from the GOES-16 ABI are available from \cite{NOAA_GOES_AWS}.
\item Observations from the Himawari 8 and 9 AHI are available from \cite{NOAA_HIMAWARI_AWS}.
\item SEVIRI observations are available from \cite{EUMETSAT_SEVIRI}.
\item ERA5 data are available from \cite{CDS_ERA5}.
\item The WegenerNet station data is available from \cite{WegenerNet2025}
\end{itemize}

\section*{Code Availability}

The \texttt{speed} package \citep{Speed2025} has been used to create the
collocation scenes and extract the training, validation, and testing data for
the SatRain dataset.

The code to access the data and evaluate retrieval results is available through
the \texttt{satrain} package \citep{SatRain2025}

\section*{Acknowledgements}

The SatRain dataset contains modified Copernicus Climate Change Service
information. Neither the European Commission nor ECMWF is responsible for any
use that may be made of the Copernicus information or data it contains.

The work of Simon Pfreundschuh work has been supported by NASA grant 80NSSC22K0604.

\section*{Author Contributions}

SP has created the dataset and drafted the manuscript. All other author's have
contributed to the design of the dataset and the writing of the manuscript.

\section*{Competing Interests}

The authors declare no competing interest.

\bibliographystyle{plainnat}
\bibliography{bibliography}
\end{document}